\documentclass[10pt,journal]{IEEEtran}

\usepackage{blindtext}
\usepackage{tcolorbox}
\usepackage{graphicx}

\usepackage{color}

\ifCLASSOPTIONcompsoc
  \usepackage[nocompress]{cite}
\else
  \usepackage{cite}
\fi
\usepackage{ulem}
  \usepackage{graphicx}

\ifCLASSINFOpdf
\else
  \usepackage{graphicx}
\fi
\ifCLASSOPTIONcompsoc
 \usepackage[caption=false,font=footnotesize,labelfont=sf,textfont=sf]{subfig}
\else
 \usepackage[caption=false,font=footnotesize]{subfig}
\fi
\usepackage{url}
\usepackage{color}
\begin{document}
\normalem
%\title{Networked Labs on a Chip (NLoC): exploiting pure hydrodynamic droplet manipulation for supporting communications and %networking in microfluidic systems}
%\title{$\mu$-NET: a microfluidic network supporting molecular computing}
\title{SD-WISE: A Software-Defined WIreless SEnsor network}


\author{Angelos-Christos~G.~Anadiotis,~\IEEEmembership{Member,~IEEE,}
	Laura Galluccio,
	Sebastiano Milardo, 
        Giacomo~Morabito,
        and~Sergio~Palazzo,~\IEEEmembership{Senior~Member,~IEEE}       
\thanks{
	A preliminary version of this paper was presented at IEEE Infocom 2015 with the title ``SDN-WISE: Design prototyping and experimentation of a stateful SDN solution for wireless sensor networks".
	\vspace{3 mm}
}
% <-this % stops a space
\IEEEcompsocitemizethanks{\IEEEcompsocthanksitem A.-C. Anadiotis is with the EPFL (work done during his time at CNIT UdR Catania).\protect\\
E-mail: angelos.anadiotis@epfl.ch
\IEEEcompsocthanksitem Laura Galluccio, Giacomo Morabito, and Sergio Palazzo are with the University of Catania.
% \protect\\
E-mail: \{name.surname\}@dieei.unict.it
\IEEEcompsocthanksitem Sebastiano Milardo is with the University of Palermo.
\protect\\
Email: sebastiano.milardo@unipa.it
}% <-this % stops an unwanted space
% \thanks{Manuscript received April 19, 2005; revised September 17, 2014.}
}
\maketitle
\thispagestyle{empty}
\pagestyle{empty}
%%%%%%%%%%%%%%%%%%%%%%%%%%%%%%%%%%%%%%%%%%%%%%%%%%%%%%%%%%%

%\vspace{-1.5 cm}


% \fi
\IEEEpeerreviewmaketitle
\begin{abstract}
SD-WISE is a complete software-defined solution for wireless sensor (and actuator) networks (WSNs).
SD-WISE has several unique features making it a flexible and expandable solution 
that can be applied in heterogeneous application domains.
Its fundamental feature is that it provides software abstractions of the nodes' 
resources on both the controller and the nodes sides.
By leveraging these abstractions, SD-WISE (i) extends the Software Defined Networking (SDN) approach to WSNs, 
introducing a more flexible way to define flows as well as the possibility to 
control the duty cycles of the node radios to increase energy efficiency;
(ii) enables network function virtualization (NFV) in WSNs;
(iii) leverages the tight interplay between trusted hardware and software to 
support \emph{regulation compliant} behavior of sensor nodes.
% In fact, SD-WISE extends the Software Defined Networking (SDN) approach to WSNs, 
% introducing a more flexible way to define flows as well as the possibility to 
% control the duty cycles of the node radios to increase energy efficiency. 
% Also, SD-WISE enables network function virtualization (NFV) in WSNs by 
% exploiting abstractions .
% Finally, SD-WISE leverages the strict interplay between trusted hardware and software to support \emph{regulation compliant} behavior of sensor nodes.
In this paper SD-WISE is introduced, its major operations are described, and its features are demonstrated in three relevant scenarios, thus assessing the effectiveness of the approach.
\end{abstract}
%%%%%%%%%%%%%%%%%%%%%%%%%%%%%%%%%%
\section{Introduction} \label{Introduction}
%%%%%%%%%%%%%%%%%%%%%%%%%%%%%%%%%%
The \emph{Software Defined Networking} paradigm is changing the way in which networks are conceived with disruptive implications on network  design, deployment, operation, and maintainance.
In the last few years, the industrial and academic communities have devoted relevant efforts to SDN development, and nowadays well established SDN solutions are available for both wired and wireless infrastructured network domains.

The adoption of SDN in wireless \emph{infrastructureless} networks and, more specifically, in wireless sensor (and actuators) networks (WSNs), is at its infancy, instead.
Accordingly, in this paper we introduce a software-defined complete solution for WSNs which we call \emph{SD-WISE}.
Major innovative features of SD-WISE are the following:
\begin{itemize}
	\item
		SD-WISE includes a software defined networking (SDN) solution for WSNs which extends the general OpenFlow approach to such domain.
		As in OpenFlow, SD-WISE nodes maintain a table (the \emph{WISE Table}) the entries of which specify how to distinguish packets belonging to certain flows (the \emph{Rules}) and  how such packets should be treated (the \emph{Actions}).
		Entries in the table are sent to the nodes by an external \emph{Controller}.
		The three major distinctive characteristics of SD-WISE as compared to OpenFlow and recent SDN proposals for WSNs are the following:
		\begin{itemize}
			\item
				{\bf It is stateful}. State information is maintained inside each sensor node and can be modified by executing the \emph{Actions}.
			\item 
				{\bf It supports a flexible definition of the Rules}. Rules can involve any portion of the packet to identify the corresponding flow and can exploit a large set of relational operators to identify whether a certain condition is satisfied or not.
			\item
				{\bf It is energy-aware}. One of the most (if not \emph{the most}) precious resources in WSNs is energy. 
				Energy consumption is often reduced in WSN by exploiting duty cycles, i.e., nodes spend large portions of time in OFF state, and transmission power control, i.e., nodes transmit at the power level which best suits the current transmission conditions. 
				The above techniques have not been considered in OpenFlow design because energy is not a major problem in infrastructured networks.
				SD-WISE instead has been designed to support both duty cycle and transmission power control.
		\end{itemize}
	\item
		SD-WISE extends network function virtualization (NFV) to WSNs. 
		To this purpose we extend the Open Networking Operating System  
(ONOS), which is currently under development for infrastructured networks, and 
exploit the capability of sensor nodes to host an (often lightweight) operating 
system. 
		For example, Raspberry Pi and other embedded devices can host operating systems like Contiki or RIoT \cite{Contiki, RIOT}.
	\item
		SD-WISE leverages strict interplay between trusted hardware and software to guarantee that nodes behavior is compliant to context-based rules.
		To this purpose SD-WISE builds on the work of the Trusted Computing Group in the context of Trusted Platform Modules (TPM) which are now commercially available for several embedded systems.
\end{itemize}

SD-WISE design is highly modular, in order to address the high variety of 
WSN devices and their capabilities as well as the large differences in the requirements stemming from the application scenarios.
In fact, given such heterogeneity there can be no 
\textit{one-size-fits-all} SD-WISE deployment.
Accordingly, different parts of the SD-WISE node or controller stacks may be 
deployed in each context.

%SD-WISE design is highly modular, in order to address the high heterogeneity of WSN devices and their capabilities. In fact, given the wide variety of WSN applications, there can be no \emph{one size fits all}. Accordingly, different parts of the SD-WISE node or controller stack may be deployed in each context.

% SD-WISE design is extremely modular, this has several advantage in general, and 
% is particularly important in our context which is characterized by high 
% heterogeneity of devices which might not be capable of implementing the entire 
% SD-WISE stack.

The objective of this paper is to present the full SD-WISE architecture and to 
demonstrate its major features in three relevant cases. 
In fact, we will show how SD-WISE can exploit an OpenFlow-like definition of forwarding rules in a real WSN testbed.
We will also demonstrate how SD-WISE network function virtualization capabilities can be exploited to support geographic routing.
Note, however, that the same approach can be applied to support any other protocol such as 6LoWPAN or ZigBee.
Finally, we will demonstrate how SD-WISE can exploit the interplay between trusted hardware and software to guarantee that certain operations cannot be performed by nodes in certain contexts which might be defined \emph{on-the-fly} in a very dynamic way.

More specifically, the rest of the paper is organized as follows.
In Section \ref{RelatedWork} we provide an overview of relevant literature.
In Sections \ref{SDWISE} and \ref{SDWISEProcedures} we present the major characteristics of SD-WISE.
Its behavior in the relevant use cases will be the subject of Section \ref{UseCases}.
Finally, in Section \ref{Conclusions} we will draw some conclusions.

%%%%%%%%%%%%%%%%%%%%%%%%%%%%%%%%%%
\section{Related work} \label{RelatedWork}
%%%%%%%%%%%%%%%%%%%%%%%%%%%%%%%%%%

SD-WISE is a holistic, integrated solution for software-defined WSNs.
Its goal is to render WSNs modular in both communication and processing, while 
enforcing regulation compliance.
To achieve this, it leverages software-defined networking, network function 
virtualization and in-device security technologies, respectively.
More specifically, SD-WISE first builds a node architecture, which supports a 
novel SDN protocol, as well as the deployment of functions, developed by users, 
inside the nodes.
The node architecture also incorporates a trusted environment, which relies on 
TPM in order to verify that the node behaves within a given regulatory context.
Then, the interaction between the applications and the nodes is performed 
through a network operating system, which provides abstractions to instantiate 
and send SDN rules, network functions, as well as context rules.

In this section, we discuss the relevant related work in the above topics. 
More specifically, we present the most common SDN solutions used for WSN, then 
the SDN controllers that have been individually developed in the context of 
fixed networks and finally, state of the art approaches for enforcing compliance 
of regulations in sensor nodes.
Furthermore, for each solution described in the following, we outline the 
major shortcoming and the corresponding enhancement provided by the use of SD-WISE.

% Software defined networking has been established in both research and  industry as the new way of managing network infrastructures. SDN decouples 
% the control from the data plane, enabling the development of custom network 
% protocols, which better fit the particular requirements of each application. In 
% this section, we present the most common SDN solutions used for WSN, then the SDN controllers that have been individually developed 
% in the context of 
% fixed networks and 
% finally, state of the art approaches for enforcing compliance of regulations in 
% sensor nodes.
% Furthermore, for each solution described in the following we will outline the major shortcoming along with the corresponding enhancement provided by SD-WISE.

\textbf{Solutions for software-defined WSN.} 
Sensor OpenFlow~\cite{6324377} is the first attempt 
to implement an SDN protocol for WSNs. It follows the OpenFlow architecture, by 
considering that the nodes should maintain a flow table with entries of 
specific, predefined format. 
Sensor OpenFlow supports in-network processing mainly to enable data 
aggregation, as commonly done in WNS for energy preservation.

Note that Sensor OpenFlow  cannot support the wide range of protocols, either standard or proprietary that 
have been proposed in the context of WSNs. 
To address the above issue, the \emph{Software Defined Wireless Network} approach (SDWN), introduced in~\cite{6385039}, leverages flexible entries in 
the flow tables maintained by the sensor nodes so achieving higher efficiency in terms of energy and communication resources. 
Moreover, SDWN supports duty cycles to save energy of the nodes.

TinySDN~\cite{7387950} focuses on the support of SDN operations across different platforms which is achieved by building on TinyOS. 
TinySDN enables interoperability of SDN-enabled nodes with several controllers, and has been implemented and tested with the Cooja simulator.

In SD-WISE we propose a stateful extension of SDWN~\cite{6385039}, which has already been shown to outperform existing WSN 
protocols in terms of delay and resource utilization efficiency \cite{7172291}.
By exploiting state information and network function virtualization SD-WISE allows nodes to take forwarding decision without querying the Controller, when local information is sufficient to the purpose. 
This results in further reduction of the delay and increased efficiency in the use of system resources.

\textbf{SDN Controllers}. Even though 
OpenFlow~\cite{McKeown:2008:OEI:1355734.1355746} is the most 
well-known SDN protocol, several efforts have already been made towards 
controlling the network routing-plane, such as 
SANE~\cite{Casado:2006:SPA:1267336.1267346}, 
ETHANE~\cite{Casado:2007:ETC:1282427.1282382}, 
4D~\cite{Greenberg:2005:CSA:1096536.1096541} and 
RCP~\cite{Caesar:2005:DIR:1251203.1251205}. On the other hand, after the wide 
adoption of OpenFlow, several controllers have been developed, such as 
NOX~\cite{Gude:2008:NTO:1384609.1384625}, 
Floodlight~\cite{floodlightweb}, and 
Beacon~\cite{Erickson:2013:BOC:2491185.2491189}.
Even though these controllers were successfully applied in the first 
days of SDN, they were centralized and therefore, have been later replaced by 
distributed solutions, which are able to scale in order to meet the 
requirements of today's large scale SDN deployments.

OpenDaylight~\cite{6918985} is a distributed network operating system, which 
has been developed as a collaborative project among several universities and 
vendors. OpenDaylight supports sensor nodes in the context of  
IoT ecosystems through the IoTDM project, and targets the integration of 
information coming from sensor nodes in the cloud. Therefore, it does not 
support full interoperability of sensor nodes and switches in the network layer.

ONIX~\cite{Koponen:2010:ODC:1924943.1924968} is a distributed SDN controller, 
which identifies the scalability constraints of the aforementioned centralized 
solutions and builds an architecture, which distributes the network management 
functionality to several instances. However, its development has been 
discontinued, whereas it has been developed mainly for data centers and it is 
closed-source, as also described in~\cite{Berde:2014:OTO:2620728.2620744}.

In the light of these issues, the Open Network Operating System (ONOS) has been 
proposed~\cite{Berde:2014:OTO:2620728.2620744}. ONOS is based on Floodlight, 
but it is distributed and provides an extensible, layered architecture, 
in order to integrate other devices and protocols, besides OpenFlow, which is 
inherently supported. In fact, ONOS has been considered in the context of this 
work, due to its scalability properties and its highly modular architecture, 
which allows the seamless integration of wireless sensor devices, as has been 
shown in~\cite{7389118}.
In this work, we extend ONOS in order to support network function
virtualization and enforce regulatory compliance in wireless sensor nodes. 

\textbf{Regulation Enforcement}. Recently a few solutions have been proposed to 
restrict the behavior of sensors (as well as smartphones) in certain contexts.
From a conceptual point of view, such solutions radically differ on the basis of the element which is in charge of transforming context information into hardware dependent settings of the sensors and actuators.

On one extreme in some solutions the device itself is responsible of context information processing and translation of high level regulations into device level configurations.
Such solutions support high level, platform-independent definition of the restrictions that must be enforced to the device behavior.
However, they require a full trusted hardware/software stack to be run by devices, which might involve high cost and processing load.
The above drawbacks make such types of solutions not appropriate in most scenarios.

In a recent Apple's patent \cite{Tiscareno2016}, for example, a use case is envisioned which implements such approach. 
In the solution proposed in \cite{Tiscareno2016} commands are 
broadcast to smartphones exploiting infrared signals in areas where recording 
is prohibited (concert halls or movie theaters, for example) to disable all 
recording functionalities.
As already mentioned, such a type of solution assumes the availability of large processing resources in the device (actually, the patent focuses on smartphones).
Furthermore, context is defined by location (besides time, obviously), whereas cases exist in which context is defined by larger set of information.

At the opposite extreme there is the approach proposed 
in~\cite{Brasser:2016:RAT:2906388.2906390}.
In this case, a node is envisioned in the infrastructure which acts as a regulation authority and transforms rules into detailed, platform dependent configurations. 
The above configurations are sent to the devices and flashed in their memory.

Note that this approach requires minimal operations to be executed by the devices. 
In fact, it only relies on trusted reading and writing in the memory of the device and trustworthy measurements from their sensors to perform context discovery.
Nevertheless, such an approach requires the definition of restrictions to be applied to IoT devices in terms of low level, platform-dependent configuration, which is difficult to be realized given the dramatic heterogeneity of the IoT landscape.
Furthermore, exchange of large amount of data is required and the flashing operation may result in long time intervals during which the device is \emph{freezed}.

As explained in the following section, SD-WISE integrates the advantages of the two approaches described above.

%%%%%%%%%%%%%%%%%%%%%%%%%%%%%%%%%%%%%%%%%%%%%%%%%%
\section{SD-WISE} \label{SDWISE}
%%%%%%%%%%%%%%%%%%%%%%%%%%%%%%%%%%%%%%%%%%%%%%%%%%
In this section we describe SD-WISE. 
More specifically, in Section \ref{SDWISEOverview} we provide an overview of the major components and their interactions.
Then, in Sections \ref{SDWISENode} and \ref{SDWISENOS} we describe the 
 SD-WISE 
node architecture and  the SD-WISE operating system architecture.
% Finally, in Section \ref{SDWISEProcedures} we describe the most relevant procedures executed to support SD-WISE functionality.

%%%%%%%%%%%%%%%%%%%%%%%%%%%%%%%%%%
\subsection{SD-WISE overview} \label{SDWISEOverview}
%%%%%%%%%%%%%%%%%%%%%%%%%%%%%%%%%%

\begin{figure}
 \centering
 \includegraphics[width=\columnwidth]{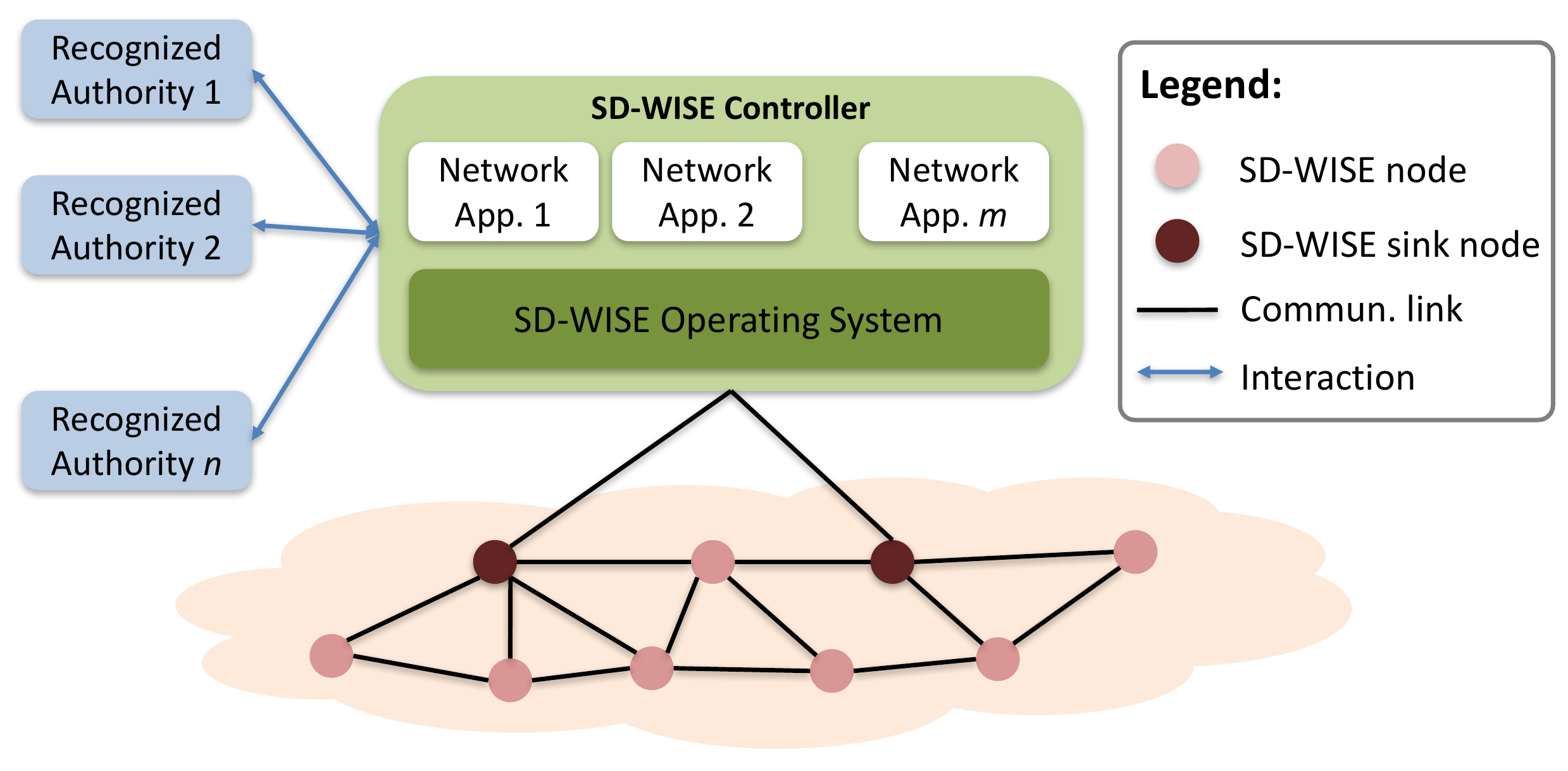}
 \caption{SD-WISE structure.}
 \label{SDWISEScenario}
\end{figure}

Operations of SD-WISE networks are  distributed between SD-WISE nodes and SD-WISE Controllers.

SD-WISE nodes are wireless sensor nodes that implement the software defined networking approach and support network function virtualization.
More specifically, in line with the mostly adopted SDN solutions, the forwarding behavior of SD-WISE nodes is determined by the content of a table which we call \emph{WISE Flow Table}. Each entry of the WISE Flow Table specifies how to treat packets with certain characteristics.
Accordingly, a node receiving a packet browses its WISE Flow Table to check 
whether an entry related to the received packet exists.
If this is the case,  the packet is treated as specified in the WISE Flow Table entry.
Otherwise, the SD-WISE node sends a request to the SD-WISE Controller and, upon receiving the response, inserts the corresponding new entry in the WISE Flow Table.
To execute such a procedure, SD-WISE nodes have to set information about a path towards the SD-WISE Controller. This is achieved by executing the Topology Discovery Protocol as specified in Section \ref{SDWISENode}.

Furthermore, SD-WISE nodes execute a software stack that provides an API 
 used to access any information related to the nodes, as well as 
download, deploy and execute application layer functions, specified in the 
SD-WISE Controller.
% Furthermore, SD-WISE nodes execute a software stack that creates abstractions of the node resources and provides the APIs required to download, deploy, and execute the application layer functions determined by the SD-WISE Controller. 
In this way SD-WISE supports the network function virtualization paradigm.

SD-WISE nodes execute a \emph{trusted} firmware that has full control on nodes' 
peripherals. 
This firmware regulates the behavior of the peripherals according to the 
directives generated by a remote \emph{Recognized Authority}, which is identified 
by the SD-WISE Controller on the base of the current node context.
To ensure that the trusted firmware is authentic, SD-WISE nodes are equipped 
with a \emph{Trusted Platform Module} (TPM) which is leveraged to execute 
software attestation.

In accordance with the current trends concerning SDN, the SD-WISE Controller is a 
software suite consisting of a network operating system (NOS), called 
\emph{SD-WISE Operating System} (or \emph{SD-WISE OS} in short), and several \emph{network applications}.
A network application defines the way the network will treat a subset of 
packet flows.
Therefore, routing protocols are implemented as network applications.
Since different network applications can be installed and run simultaneously by 
the same Controller, it becomes simple to provide the most suitable treatment to different applications 
with heterogeneous characteristics and needs.

SD-WISE OS provides a rich set of APIs that allow network applications to manage different types of remote devices using unified 
abstractions to represent them as well as to create and send rules to them.
% On the northbound the SD-WISE Operating System offers well defined APIs 
% enabling Network Applications to obtain controlled access to the Device 
% Management modules of the SD-WISE nodes.
In fact, the SD-WISE OS transforms the policies set by network 
applications into directives that determine the way SD-WISE nodes will generate 
and treat the relevant packets.
As an example, in Section \ref{GeographicRouting} we will discuss how a network 
application can implement Geographic Routing by exploiting the APIs of the 
SD-WISE Operating System.

The SD-WISE Controller is also responsible for acting as an intermediary 
between the SD-WISE nodes and \emph{Recognized Authorities}, i.e., entities that 
have the power to limit the operations of nodes based on regulations or other 
policies, as previously explained.

%%%%%%%%%%%%%%%%%%%%%%%%%%%%%%%%%%
\subsection{SD-WISE node architecture} \label{SDWISENode}
%%%%%%%%%%%%%%%%%%%%%%%%%%%%%%%%%%

\begin{figure}
 \centering
 \includegraphics[width=\columnwidth]{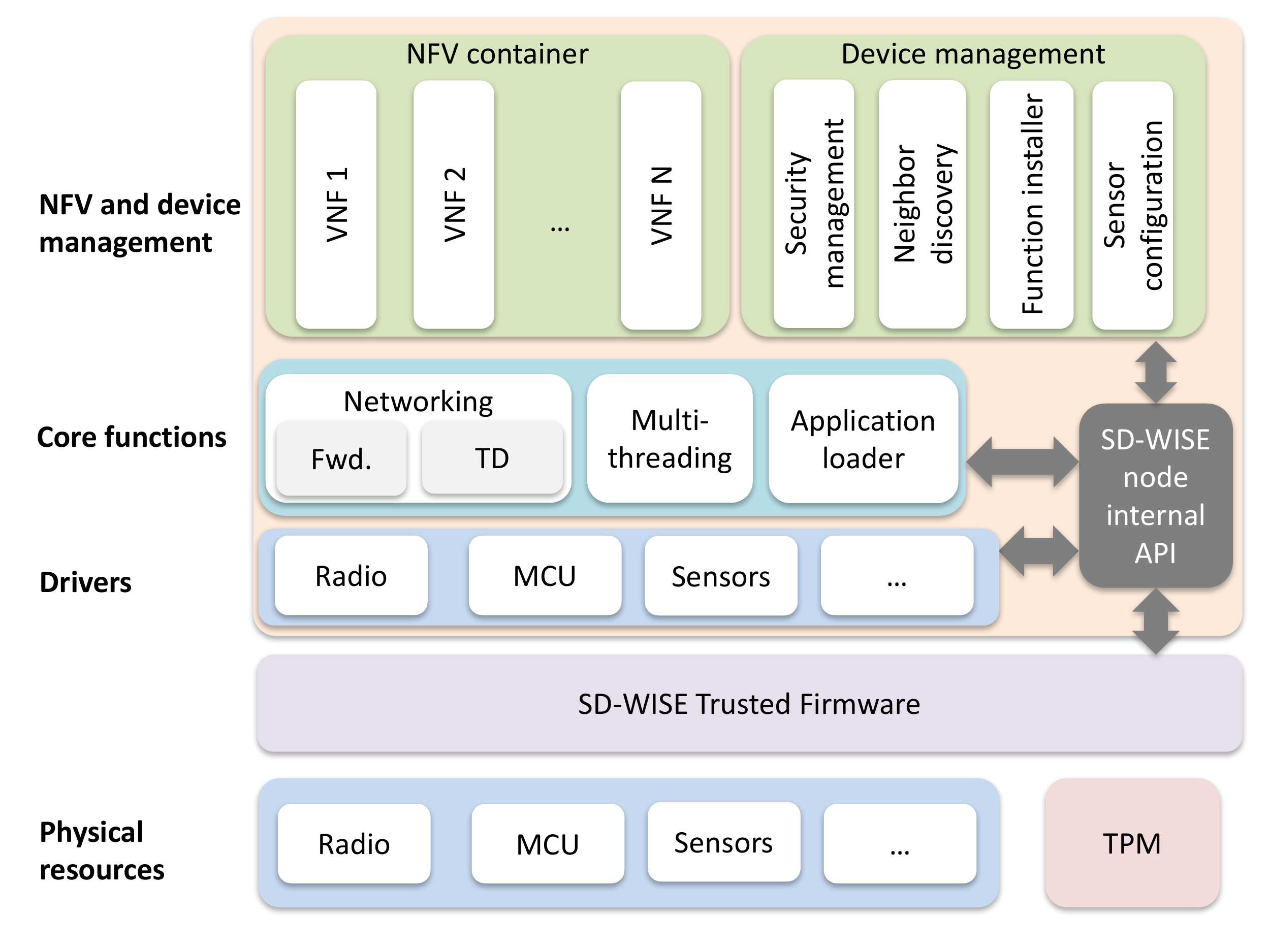}
 \caption{SD-WISE node architecture.}
 \label{fig:SDWISENodeArchitecture}
\end{figure}

The architecture of  SD-WISE nodes is based on the typical approach of the operating systems for sensor nodes such as Contiki and RIOT, and is represented in Figure \ref{fig:SDWISENodeArchitecture}.

As regards the hardware characteristics, SD-WISE nodes have a \emph{Trusted Platform Module} (TPM) besides the usual resources such as radio transceiver, MCU, sensors, actuators, memory, etc.

% Seb: I'm adding some details on the TPM as it is not clear what is the role of the TPM and how by leveraging a TPM we implement trust
The TPM implements in hardware four basic security primitives:

\begin{itemize}
\item Random number generation;
\item Cryptographic key generation and storage;
\item Data Encryption/Decryption;
\item Hashing;
\end{itemize}

These primitives are leveraged by SD-WISE to authenticate trusted elements, to ensure that the software run by a device is trusted, and that the values provided by sensors are authentic and have been generated and processed by trusted sensors ~\cite{tpm_spec}.

On top of the hardware the {\bf SD-WISE Trusted Firmware} is executed, which has full control on the hardware resources.
By exploiting the TPM functionality, the SD-WISE Trusted Firmware can be authenticated both locally and remotely by executing software integrity measurement and attestation, respectively.
Note that the SD-WISE Trusted Firmware is in the position to filter all 
interactions to/from the hardware and therefore, has full control on all 
peripherals - sensors and actuators.

The \textbf{Drivers} implement the core abstractions of the hardware resources 
of each node.
Access to the physical resources is only given through the SD-WISE Trusted 
Firmware.

% {\bf Drivers} offering an abstraction of the hardware resources have access to 
% the physical resources through the SD-WISE Trusted Firmware.

% The abstractions of the SD-WISE node resources are provided by the SD-WISE 
% internal API and implemented by the Drivers are offered to the {\bf Core} 
% layer.
The \textbf{Core functions} layer builds on top of the SD-WISE node 
abstractions and provides key functionalities such as the support of 
Networking, Multithreading  and 
Application Loading, which allows to load and execute new applications in the 
SD-WISE node without the need to restart the node. 
Recent versions of the major operating systems for embedded systems, such as 
Contiki and RIOT, offer this functionality.

The Networking component of the Core layer implements two fundamental 
protocols: the \emph{SD-WISE Forwarding Protocol} and the \emph{Topology 
Discovery Protocol}.
The SD-WISE Forwarding Protocol is mostly responsible for the management of 
incoming packets with an approach which is derived from OpenFlow and will be 
described in Section \ref{SDWISEProcedures}.
The Topology Discovery (TD) protocol, instead, is executed by SD-WISE nodes for 
generating local topology information and delivering it to the Controller. 
More specifically, the TD protocol maintains updated information about the next 
hop of each node towards the Controller as well as its current neighbors. 
To this purpose all sinks in the SD-WISE network periodically and (almost) 
simultaneously transmit a Topology Discovery packet (TD packet) over the 
broadcast wireless channel. 
This packet contains the identity of the sink that generated it, the battery 
level of each node transmitting it, and the current distance from the sink 
which is initially set to 0.
A node $A$, upon receiving a TD packet from node $B$ (note that B can be also a sink), 
performs the following operations:
\begin{enumerate}
	\item
		inserts $B$ in the list of its current neighbors along with the current RSSI and the battery level.
		Obviously, if $B$ is already present in the list of current neighbors, then only the RSSI and battery level values are updated;
	\item
		controls whether it has recently received a TD packet with a lower value of the current distance from the sink.
		If this is not the case, then node $A$ updates the value reported in the TD packet to the current value plus one and sets its next hop towards the Controllers equal to $B$;
	\item
		sets its battery level in the corresponding field of the TD packet;
	\item
		transmits the updated TD packet over the broadcast wireless channel.
\end{enumerate}
Periodically, each node generates a packet containing its current list of neighbors and sends it to the Controller.
Note that the list of neighbors is periodically cleared.
Nodes receiving packets directed towards the Controllers relay them to the node 
that is closer (in terms of number of hops) to the sink.

The {\bf Network function virtualization and device management} layer runs on top of the Core layer.
At this layer the {\bf virtual network functions}, which can be loaded \emph{on 
the fly}, are executed.
The network functions make use of the Core API, which gives them access to the 
node resources.
Furthermore, at this layer node management functions run. 
Examples of such functions include node configuration, security management, and applications/NF installation.

Finally, interactions between all SD-WISE node components occur through the 
{\bf SD-WISE node internal API}. 

%%%%%%%%%%%%%%%%%%%%%%%%%%%%%%%%%%
\subsection{SD-WISE operating system architecture} \label{SDWISENOS}
%%%%%%%%%%%%%%%%%%%%%%%%%%%%%%%%%%

\begin{figure}
 \centering
 \includegraphics[width=\columnwidth]{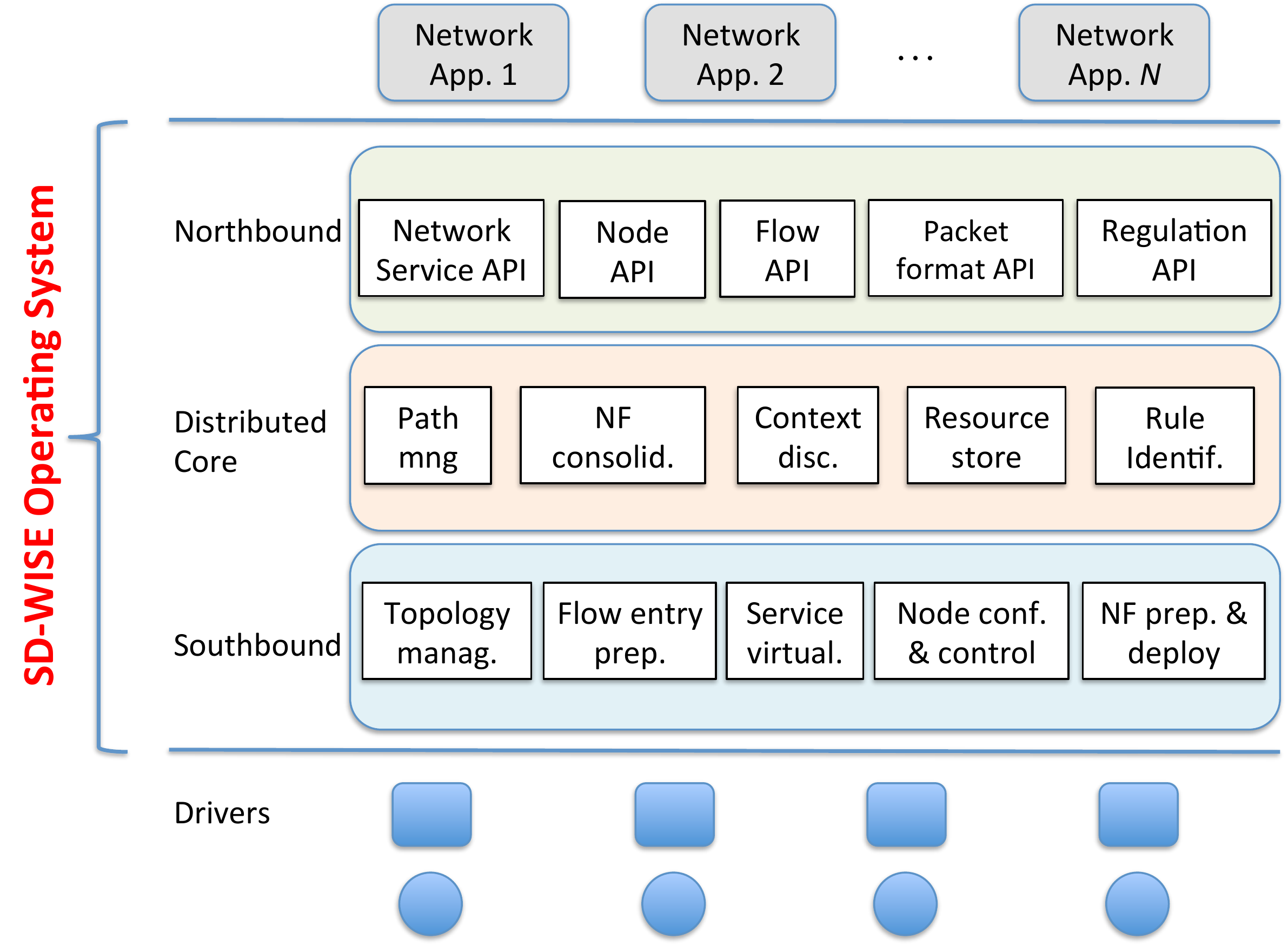}
 \caption{SD-WISE Operating System architecture.}
 \label{SDWISEOSArchitecture}
\end{figure}

The architecture of the SD-WISE Operating System extends the one of the Open Network Operating System (ONOS) and is depicted in Figure \ref{SDWISEOSArchitecture}.

Like ONOS, the SD-WISE Operating System consists of three layers called \emph{Southbound}, \emph{Distributed Core}, and \emph{Northbound}.

Objective of the {\bf Southbound} is to provide the higher layers with services supported by the   existing resources which are independent of 
platform-specific features, such as the layer 2 packet header format, the 
addressing scheme, the sensor data format, etc.
Accordingly, major responsibilities of the Southbound are the creation and 
management of the network topology, the formatting and management of the flow 
entries that will be sent to the SD-WISE nodes, the virtualization of the 
services offered by the sensor and actuators deployed in the nodes, the 
configuration and control of the SD-WISE nodes, and the preparation and 
deployment in the nodes of the network functions (NF).

The {\bf Distributed Core} is responsible of most critical management functions. 
For example, it is responsible of identifying optimal communication paths and transform them into sequences of WISE table entries which will be deployed in the SD-WISE nodes.
Furthermore the Distributed Core is responsible of the so-called {\it Network Function consolidation}, i.e., identify the subset of nodes where NF should be installed at the Device management and applications layer, as described in the previous Section \ref{SDWISENode}.
The Distributed Core implements the functionality needed for discovering the context in which each SD-WISE node is operating and identify the respective {\it Recognized Authority}.
The distributed core maintains the so called \emph{Resource Store} updated , i.e., a database containing information and characterization of the resources provided by the active SD-WISE nodes.
The Resource Store can be queried by the applications through the Northbound.
Finally, the identification of the rules that SD-WISE nodes must comply with, based on the high level directives coming from the relevant Recognized Authorities is also within the scope of the Distributed Core layer.

The {\bf Northbound} is responsible of providing applications with access to the services offered by SD-WISE networks.
Such access must be completely independent of the specific characteristics of the underlying physical resources.
More specifically, the Northbound specifies the high level features of the services (i.e., unicast, broadcast, anycast, geocast, etc.) offered by the WSN.
In the Northbound, the abstractions of SD-WISE nodes are also provided. 
In this context, we extend what is available in ONOS, by introducing node features that are specific of sensor networks and cannot be found in traditional infrastructured networks. 
Examples include abstractions of the sensors and of the actuators deployed in the node, the level of battery, the current state\footnote{Recall that sensor nodes spend most of the time in \emph{idle} state to reduce energy consumption.}.
We extend ONOS for the Flow API as well. 
In fact in SD-WISE it is possible to define flows utilizing several relational operators, as anticipated in Section \ref{Introduction} and discussed deeply in the following Section \ref{SDWISEProcedures}.
The Packet API, instead, is exactly the same as in ONOS.
Finally, for what concerns the communication involving the most external layers of the proposed architecture,  while on the one hand the Northbound will define the interface that Recognized Authorities must use to define context-based rules in platform-independent way, on the other hand, the Southbound will interact with the physical devices enforcing secure and authenticated sessions.  
It is worth noting that in many cases IoT devices may lack of direct Internet connectivity or may have limited resources in terms of memory or computational power, therefore identity validation using standard protocols like X.509 may be unfeasible. 
In this case, the TPM will support such validation by allowing the exchange of signed messages using pre-shared pairs of public/private keys stored inside the TPM itself. 

%%%%%%%%%%%%%%%%%%%%%%%%%%%%%%%%%%%%%%%%%%%%%%%%%%
\section{SD-WISE operations} \label{SDWISEProcedures}
%%%%%%%%%%%%%%%%%%%%%%%%%%%%%%%%%%%%%%%%%%%%%%%%%%
In this section we will describe the two major novel operations introduced by SD-WISE, that is, the \emph{SD-WISE Forwarding} included in the Networking module of the Core layer of the SD-WISE nodes and the procedures run to guarantee that SD-WISE nodes are compliant to context-based rules set by a recognized authority. 
The above operations will be detailed in  Sections \ref{SDWISEForwarding} and \ref{RegulationCompliance}, respectively.

%%%%%%%%%%%%%%%%%%%%%%%%%%%%%%%%%%
\subsection{SD-WISE forwarding} \label{SDWISEForwarding}
%%%%%%%%%%%%%%%%%%%%%%%%%%%%%%%%%%

For what concerns forwarding, the behavior of SD-WISE nodes is completely encoded in three data structures, namely: the \emph{WISE State}, the \emph{Accepted IDs Array}, and the \emph{WISE Flow Table}.
Along with most SDN approaches, such structures are filled with the information coming from the Controller, running in appropriate server.
In this way the Controller defines the networking policies which will be implemented by the SD-WISE nodes.

At any time SD-WISE nodes are characterized by the current WISE State which is an array of strings of $s_\mathrm{State}$ bits each.
The State can be modified by the Controller or by nodes themselves.

Given the broadcast nature of the wireless medium, in general sensor nodes can receive packets which are not meant for them (not even for forwarding).
The Accepted IDs Array allows each WISE node to select only the packets  which it must further process.
In fact, the header of the packets contains a field in which an ID is specified.
A node, upon receiving a packet, controls whether the ID contained in such field is listed in its Accepted IDs Array.
If this is the case, the node will further process the packet; otherwise it will drop it.
Note that SD-WISE specifies the packet format proposed in ~\cite{6385039} in which the ID field replaces the \emph{next hop address}.
Each network application can, however, override such format and specify its own through the Packet format API provided by the Northbound of the WISE Operating System as described in the previous Section \ref{SDWISENOS}.
If this is the case, a field in the packet header is used to identify the application that has generated the packet.
Nevertheless, the ID field must remain as it is required to enable SD-WISE operation over network segments in which all communications are broadcast.

In the case the packet must be processed, the sensor node will browse the entries of its WISE Flow Table.
Each entry of the WISE Flow Table contains a \emph{Matching Rules} section which specifies the conditions under which the entry applies.
Matching Rules may consider any portion of the current packet as well as any bit of the current state.
If the Matching Rules are satisfied, then the sensor node will perform an 
\emph{Action} specified in the remaining section of the WISE Flow Table entry.
Note that such action may refer to how to handle the packet as well as how to modify the current state of the node.

If no entry is listed in the WISE Flow Table whose Matching Rules apply to the current packet/state, then a request is sent to the Controller.

In order to contact the Controller, a node needs to have a WISE Flow Table entry indicating its best next hop towards one of the sinks. 
This entry is different from the others because it is not set by a Controller but is discovered by each node using the Topology Discovery protocol briefly described in Section \ref{SDWISENode}.

Note that sensor nodes have limited capabilities in terms of memory, therefore, selection of the size of the different data structures is very important. 
The optimal choice of such size depends on several deployment specific features set by the SD-WISE Operating System during the initialization phase.

%%%%%%%%%%%%%%%%%%%%%%%%%%%%%%%%%%
\subsection{Context-based regulation compliance} \label{RegulationCompliance}
%%%%%%%%%%%%%%%%%%%%%%%%%%%%%%%%%%

\begin{figure}
\centering
\includegraphics[width=3.2in]{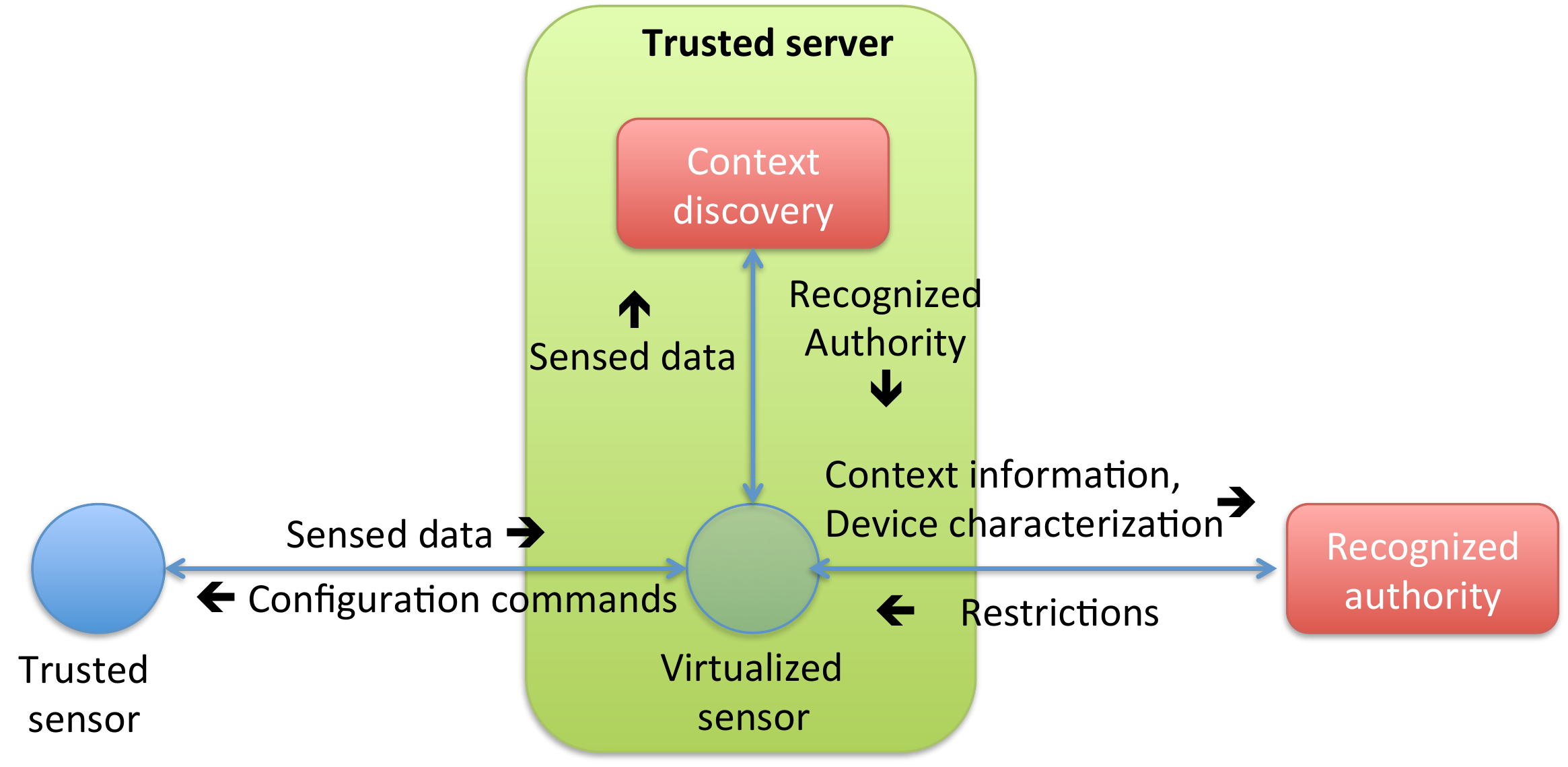}
\caption{Scheme of the operations performed by SD-WISE to guarantee regulation-compliant behavior of sensor (and actuator) nodes.}
\label{InteractionModel}
\end{figure}

The scheme of the operations performed by SD-WISE to guarantee 
regulation-compliant behavior of sensors (and actuators) is shown in Figure 
\ref{InteractionModel}.

We assume that each sensor node has a cyber counterpart, called \emph{virtual sensor} instantiated in the SD-WISE OS.
The virtual sensor acts as a proxy between the physical device and the Recognized Authority.
This approach is common in several solutions \cite{Truong2015} because exploitation of cyber counterparts of physical sensor nodes has many advantages. 
In fact, on a first hand virtual sensors are persistent, always-on entities whereas physical devices might spend large portions of the time in idle mode. 
Furthermore, virtual sensors can rely on theoretically unlimited amount of processing and communication resources, whereas physical devices might be characterized by strict resource limitations. 
Finally, the virtual sensors can define a high level abstraction of the sensor node hiding the specific details of the hardware platform implementing the physical device.
Therefore, the interactions between the sensors and the Recognized Authority are platform independent, whereas the interactions between the physical sensor and the virtual sensor can occur according to proprietary protocols. These can be optimized according to the specific needs arising from the hardware characteristics as well as the deployment scenario.

For what concerns the focus of our work, the virtual sensor will receive the information from its physical counterpart that will forward it to the \emph{Context Discovery} module to determine the current \emph{context}. 
According to \cite{Schilit1994}, the \emph{context} is defined by the identity of the node itself and the assets in its neighborhood and, therefore, relevant information is \cite{Abowd2001}:
\begin{itemize}
	\item
		\emph{Social environment}: location, nearby people, situation
	\item
		\emph{Computing environment}: nearby sensors/actuators, connectivity options
	\item
		\emph{Physical environmet}: noise, temperature, humidity, lighting, etc.
\end{itemize}
Accordingly, the values that the sensor node will collect and send to its cyber counterpart are the values measured by its sensors and the list of sensor nodes and access points in the neighborhood.
By leveraging \emph{sensor attestation} as described in \cite{Liu2012}, values received by the virtual sensor and forwarded to the Context Discovery module can be considered authentic.

In fact, the Context Discovery module uses data received by all virtual sensors to infer the context and identify the corresponding \emph{Recognized Authority}.
This is responsible for identifying the rules which sensors must comply with.
In our prototype implementation the context is defined by the position of the sensor and its owner; however, more complex cases can be easily thought.

The restrictions are sent to the virtual sensor that translates them into configuration parameters of sensors (and actuators).
Such configurations are finally transmitted to the physical sensor which will implement them and send a confirmation to its cyber counterpart.
Also in this case sensor attestation\footnote{In this case it would be more correct to call it \emph{sensor/actuator attestation}.} can be exploited to ensure that the sensor has applied the restrictions sent by its virtual counterpart.

%%%%%%%%%%%%%%%%%%%%%%%%%%%%%%%%%%%%%%%%%%%%%%%%%%
\section{SD-WISE in action} \label{UseCases}
%%%%%%%%%%%%%%%%%%%%%%%%%%%%%%%%%%%%%%%%%%%%%%%%%%
Objective of this section is to demonstrate the specific features of SD-WISE in relevant use cases.
More specifically, in Section \ref{PerformanceEvaluation} we will show how SD-WISE forwarding based on the use of the WISE Flow Table performs in a physical testbed.
Then, in Section \ref{GeographicRouting} we will provide an example of network 
application running on top of the SD-WISE OS which implements geographic 
routing and leverages the NFV capabilities of SD-WISE.
Finally, in Section \ref{ContextBasedFencing} we will present a use case in which the behavior of sensor nodes is regulated by a Recognized Authority.

%%%%%%%%%%%%%%%%%%%%%%%%%%%%%%%%%%
\subsection{WISE Flow Table-based forwarding} \label{PerformanceEvaluation}
%%%%%%%%%%%%%%%%%%%%%%%%%%%%%%%%%%

Similarly to OpenFlow, the main communication overhead of SD-WISE is represented by the exchange of information between the OS and the devices.
To measure such overhead, the performance of SD-WISE have been tested in a real testbed made of 5 wireless sensor nodes and a sink physically connected to the SD-WISE OS.

In each measurement campaign 5000 data packets have been sent, each every 15 seconds.
Different payload sizes have been considered for such packets (10, 20 and 30 bytes).
Furthermore, the time interval, $T$, between two consecutive generations of the TD packets has been changed. 
In each campaign we have set the time interval between the transmissions of local topology information to twice the value of $T$ in order to receive at least one beacon packet.

\begin{figure}
\centering
\includegraphics[width=3.5in]{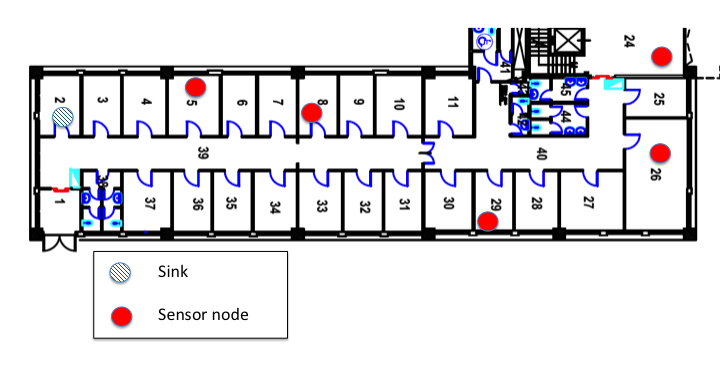}
\caption{Nodes deployment.}
\label{NodesInTheMap}
\end{figure}

\begin{figure*}
 \centering
 \subfloat[Number of hops = 3.] {
  \label{Unicast3hop}
  \includegraphics[width=0.45\textwidth]{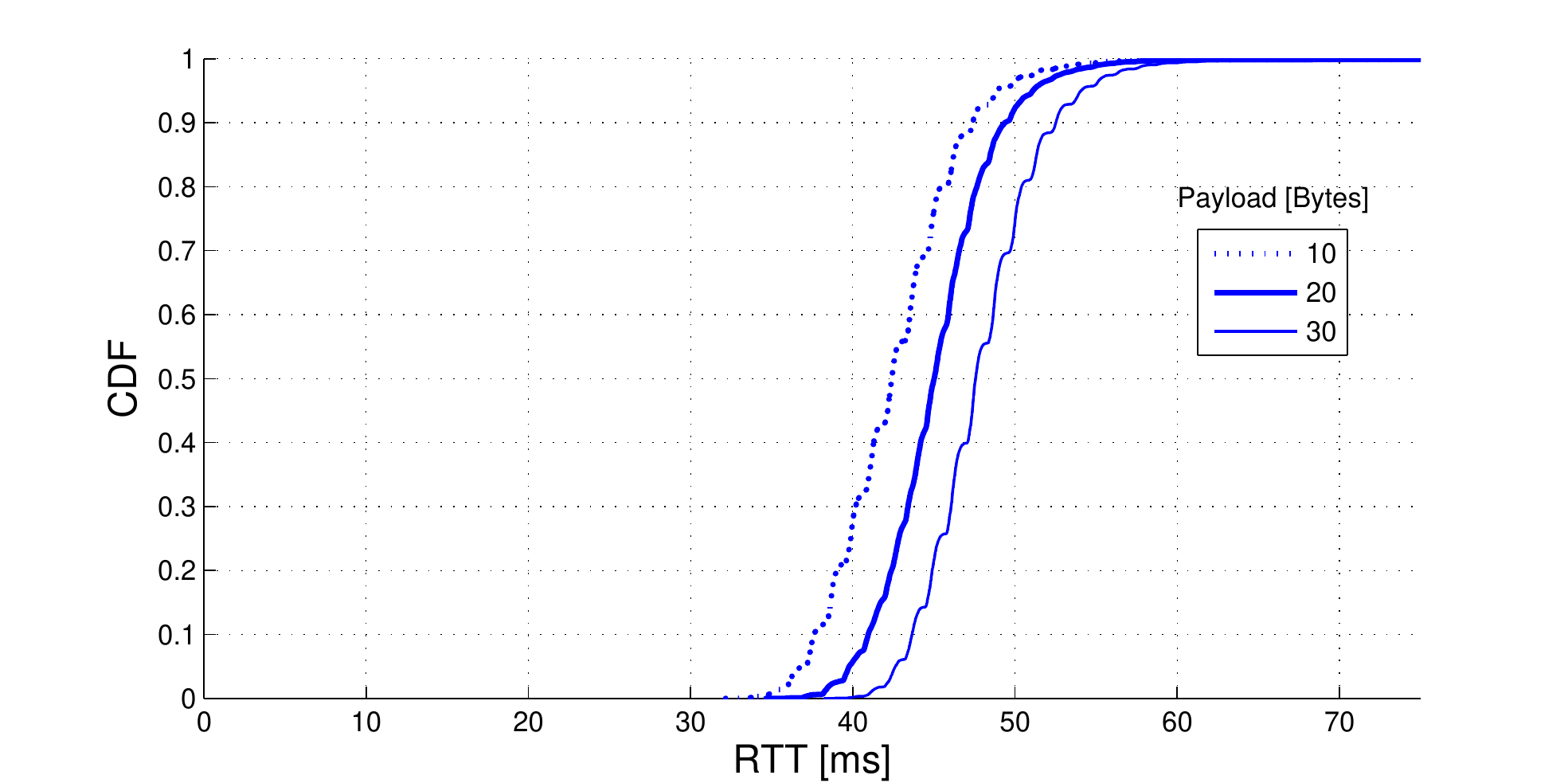}
 }\qquad
 \subfloat[Number of hops = 5.] {
  \label{Unicast5hop}
  \includegraphics[width=0.45\textwidth]{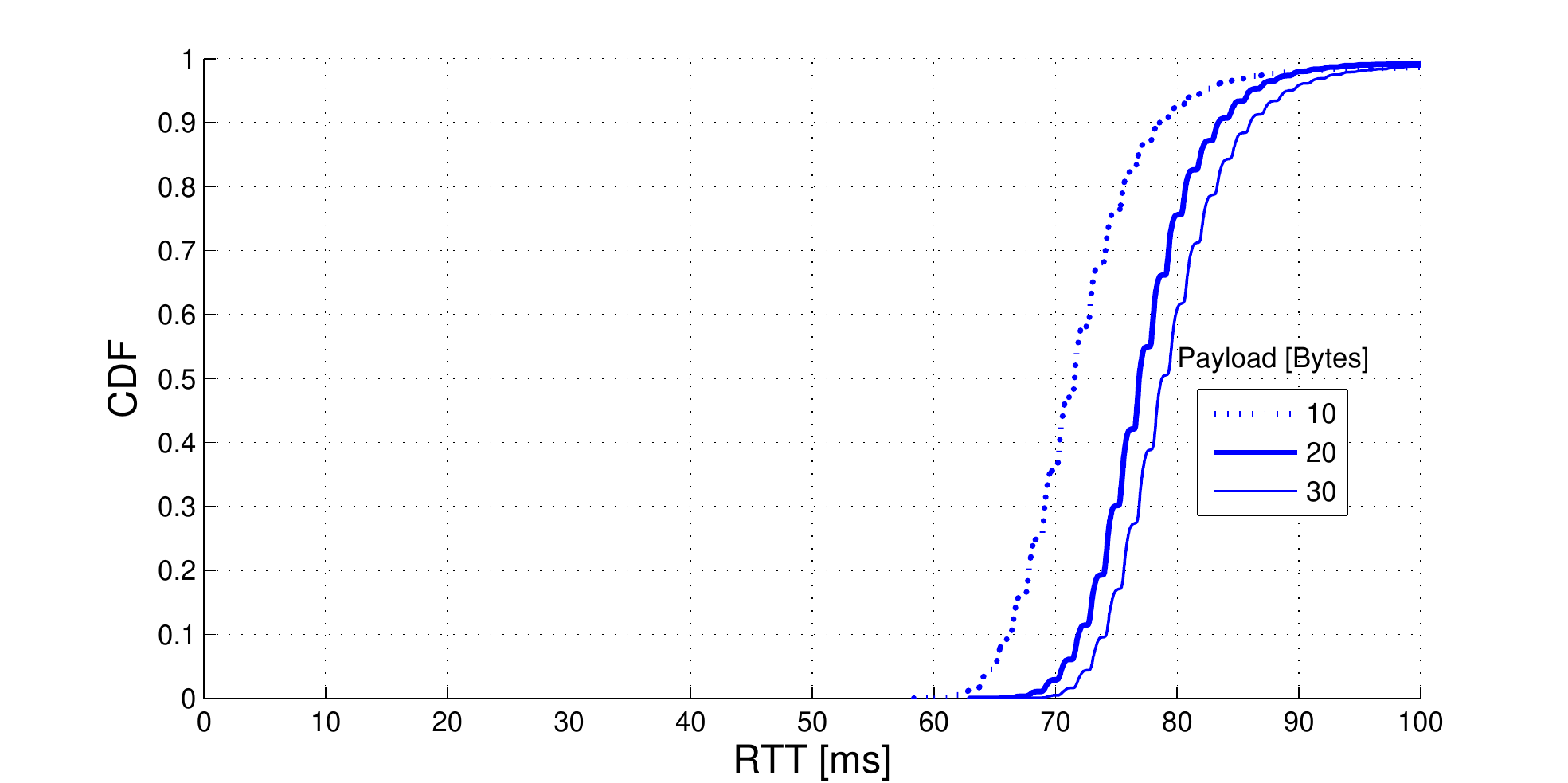}
 }
 \caption{CDFs of the RTT for different payload sizes and different distances between the source and destination node.}
 \label{RTTUnicast}
\end{figure*}

In the following we illustrate the performance achieved by SD-WISE in terms of:
\begin{itemize}
	\item
		Round Trip Time (RTT), that is, the time interval between the generation of a data packet and the reception of the corresponding acknowledgment;
	\item
		Efficiency, measured as the ratio between the number of payload bytes received by the intended destinations and the overall number of bytes circulating in the network;
	\item
		Controller response time, measured as the interval between when the Controller receives a request for a new entry and the time instant when the Controller sends the corresponding entry.
\end{itemize}

%%%%%%%%%%%%%%%%%%%%%%%%%%%%%

\begin{figure}[!t]
\centering
\includegraphics[width=3.5in]{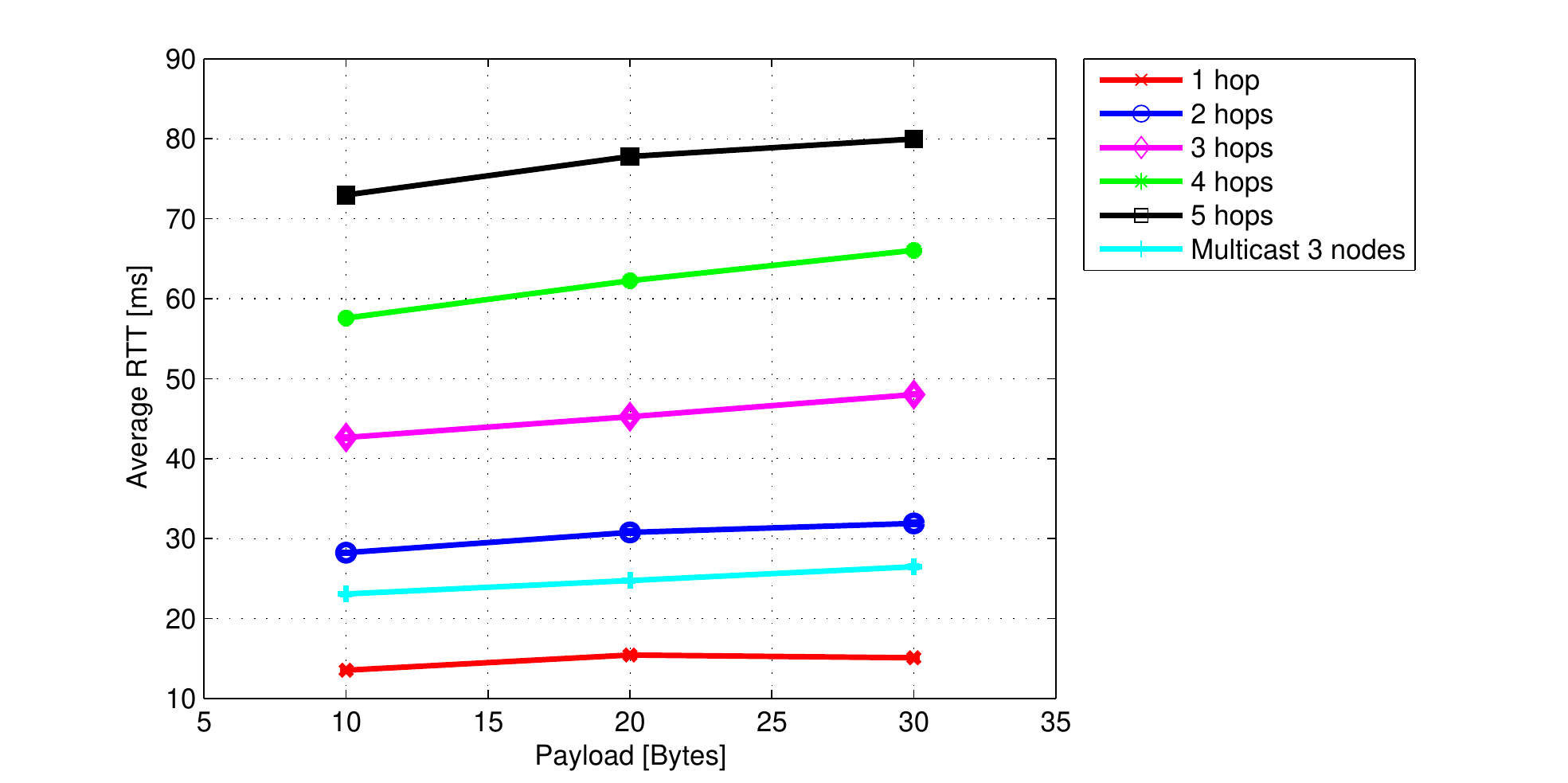}
\caption{Average RTT vs. the payload size, for different values of the number of hops.}
\label{confRTTpay}
\end{figure}

\begin{figure}[!t]
\centering
\includegraphics[width=3.5in]{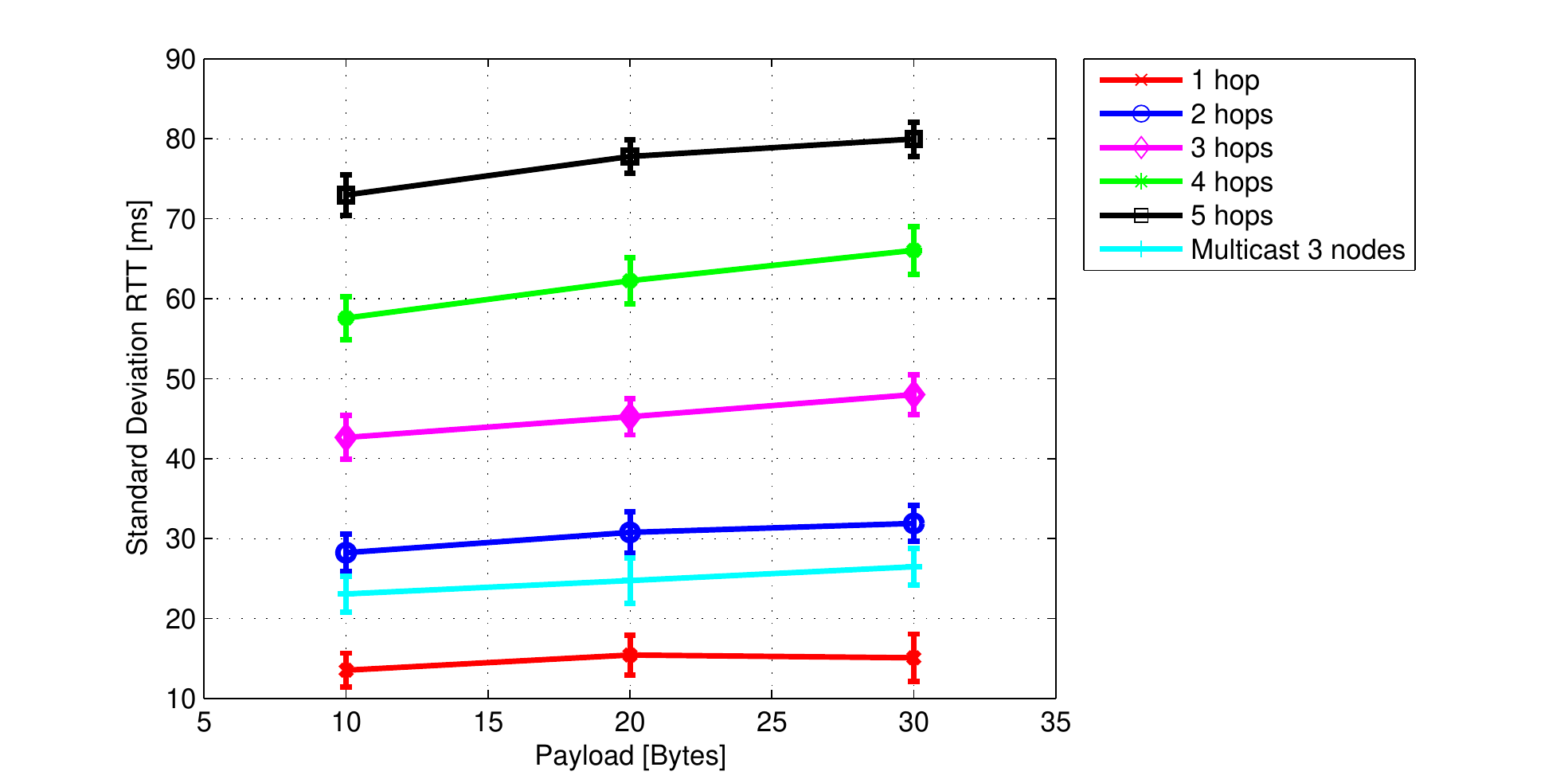}
\caption{Standard deviation of the RTT values vs. the payload size, for different values of the number of hops.}
\label{StdRTTpay}
\end{figure}

In Figures (\ref{Unicast3hop}) and (\ref{Unicast5hop}) we represent the \emph{Cumulative Distribution Functions} (CDF) of the RTT when the distance between the packet source and the packet destination is equal to 3 and 5 hops, respectively.
In each figure we represent three curves obtained for different values of the payload size (10, 20, and 30 bytes).
As expected, RTT increases as the distance and the payload increase.
Furthermore, we expect a similar behavior for the standard deviation.
Indeed, this is reflected in Figures \ref{confRTTpay} and \ref{StdRTTpay} where we show the average and the standard deviation of the RTT vs. the payload size for different values of the distance between source and destination.

\begin{figure}[!t]
\centering
\includegraphics[width=3.5in]{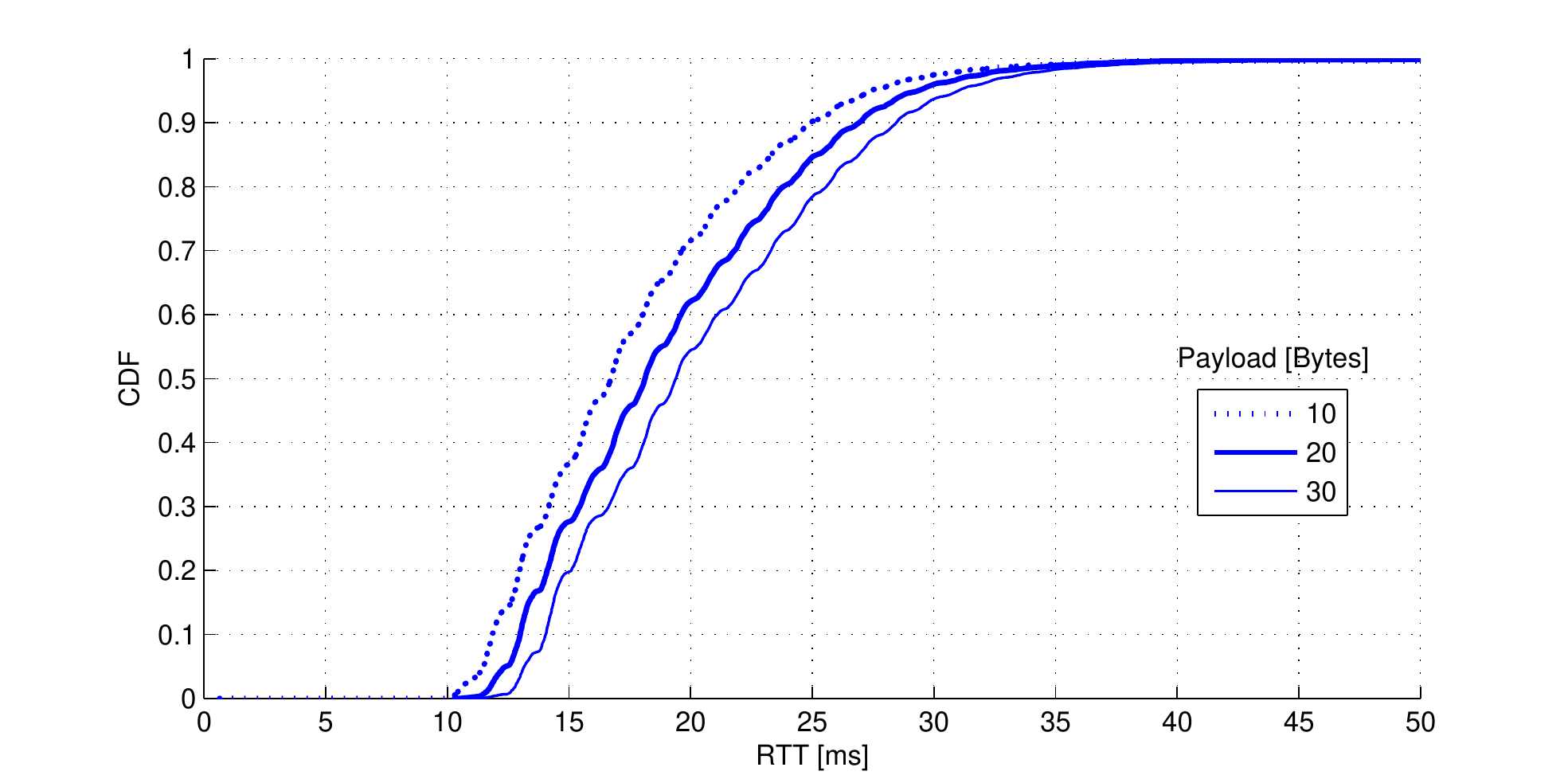}
\caption{CDF of the RTT in the multicast case for different payload sizes.}
\label{Multicast}
\end{figure}

In Figures \ref{confRTTpay} and \ref{StdRTTpay} we plot a curve for the multicast case, as well.
This has been obtained by measuring the time interval between the transmission of a packet and the reception of the acknowledgement from the last destination.
In this case, only three destinations were considered and were deployed within the radio range of the source.
Obviously, the average and the standard deviations of the RTT is slightly higher than in the analogous (one hop) unicast case.
The corresponding CDFs are represented in Figure \ref{Multicast}.

\begin{figure}[!t]
\centering
\includegraphics[width=3.5in]{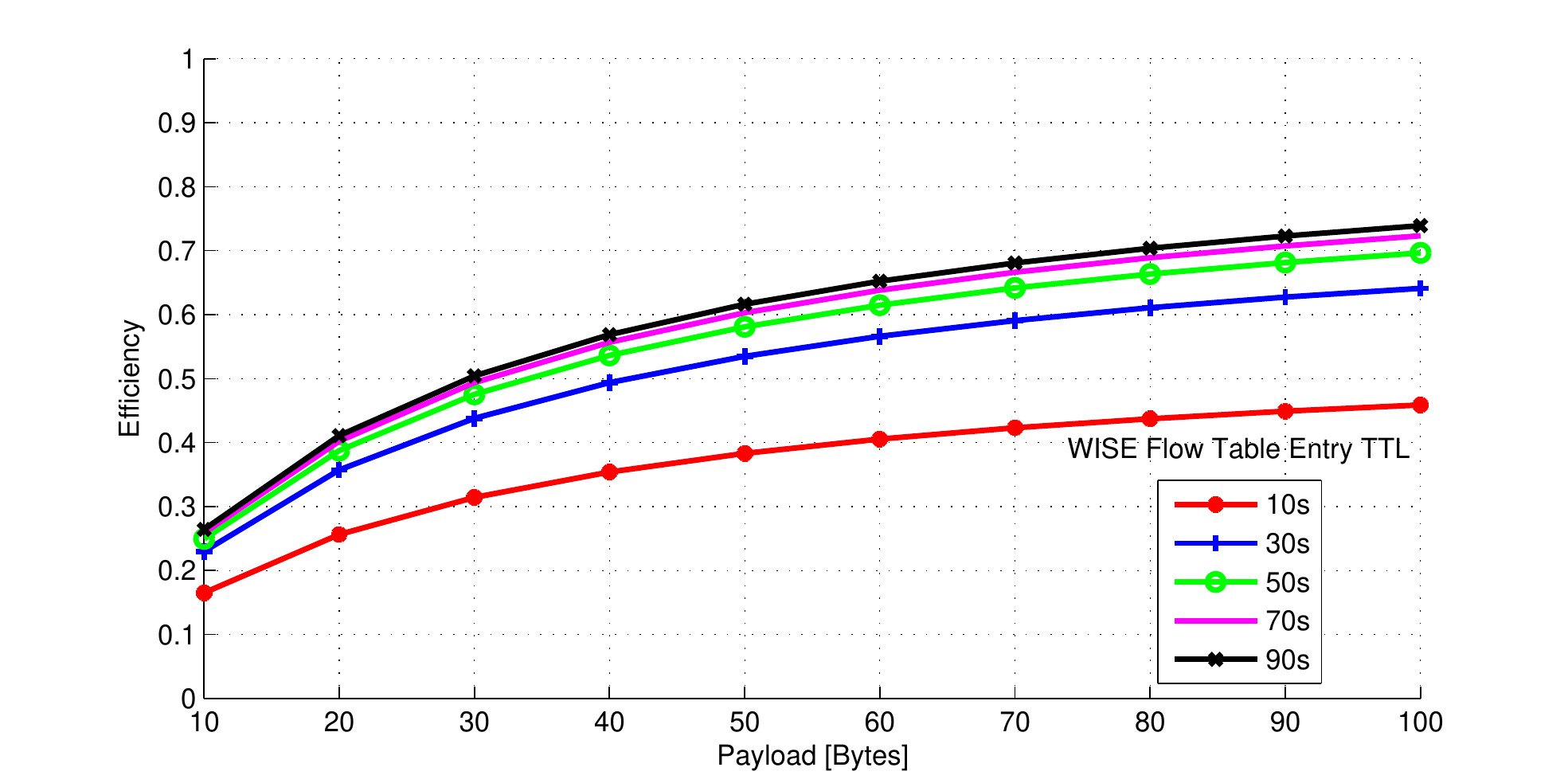}
\caption{Efficiency for different values of maximum WISE Flow Table entry TTL.}
\label{effTTL}
\end{figure}

\begin{figure}[!t]
\centering
\includegraphics[width=3.5in]{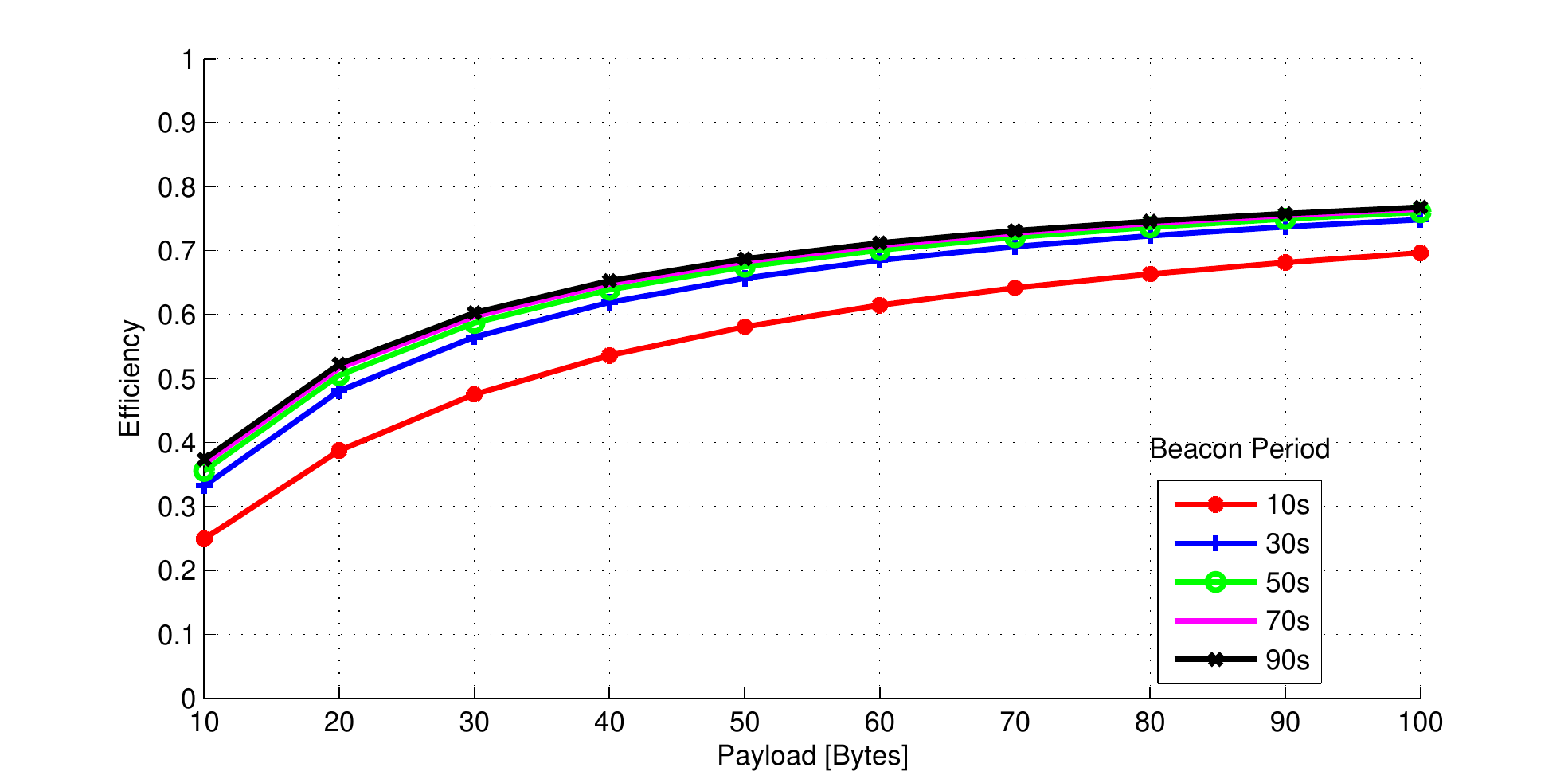}
\caption{Efficiency for different values of beacon sending period.}
\label{effBeacon}
\end{figure}

Finally, the performance in terms of efficiency are shown in Figures \ref{effTTL} and \ref{effBeacon}.
More specifically, in Figure \ref{effTTL} we represent the efficiency vs. the payload size for different values of the lifetime of an entry in the WISE Flow Table, which we denote here as TTL. Instead, in Figure \ref{effBeacon} we show the same curves obtained for different values of the interval between consecutive transmissions of the TD packets, $T$.

Note that most of the inefficiency is due to the high ratio between the header size and the payload size.

%%%%%%%%%%%%%%%%%%%%%%%%%%%%%%%%%%
\subsection{Geographic routing} \label{GeographicRouting}
%%%%%%%%%%%%%%%%%%%%%%%%%%%%%%%%%%

% Objective of this section is to show how SD-WISE can support geographic routing as a network application.
% Note that with a similar approach any other solution proposed in the literature at network layer and above can be supported by SD-WISE.
% 
% A node applying geographic routing relays an incoming packet to its neighbor which is nearest to the destination.
% To do so, therefore, it only needs to know the position of the destination, which is carried in the packet, and the positions of its immediate neighbors.
% Geographic routing has shown to be very efficient in several WSN scenarios \cite{Zorzi, MACRO, GEM}.

Objective of this section is to show how the typical, centralized, SDN 
operations shown in the previous sections, can be distributed by leveraging NFV.
More specifically, in this section we will consider \emph{geographic routing} as a proof-of-concept of network function.

In a typical SDN scenario, nodes ask the Controller to provide the rules to forward a packet to the destination.
Even though SDN-based approaches have been proven to be more efficient than the 
common distributed protocols in WSNs \cite{7172291}, they still require that 
the nodes contact the Controller when they have to send a packet to an unknown 
destination, whereas they need to maintain an entry per destination in their 
flow tables.
As it will be shown later in this section, by applying a geographic routing approach, nodes can reduce the signaling overhead and the number of flow entries that they have to keep, resulting in a significant reduction in their overall energy consumption.

Geographic routing has been shown to be very efficient in several WSN scenarios, both for unicast and multicast communications~\cite{1255648}~\cite{1498457}~\cite{6407472}.
In geographic routing, a node relays incoming packets to its immediate neighbor, which has the shortest Euclidean distance to the destination.
In order to do so, it only needs to know the position of itself, its immediate neighbors and the destination.
This information is received by the Controller, which maintains a consistent view of the network topology, by inferring the positions of the nodes from the Received Signal Strength Indication (RSSI) values that the nodes are anyway reporting periodically to the Controller.

More specifically, as explained earlier in Section \ref{SDWISENode}, in 
the context of the Topology Discovery protocol, every node periodically sends 
to the Controller a packet, which contains its immediate neighbors and the RSSI 
value corresponding to each one of them.
The typical Controller operation in this case is to update the topology graph, in order to keep a consistent view of the network topology.
In case geographic routing is enabled, the Controller also employs a localization algorithm to extract the coordinates of the nodes based on the RSSI values.
Note that there are several localization algorithms~\cite{Bachrach2005}~\cite{Mao20072529}, which provide reliable results~\cite{5972367}.
In case the position of a node has changed, the Controller sends the new coordinates to that particular node as well as its immediate neighbors.
This way, nodes are always up-to-date about the positions of themselves and their immediate neighbors.
Then, when a node wants to send a packet with geographic routing enabled, it only sends a request to the Controller for the coordinates of the packet destination and then all forwarding decisions are performed independently by each individual node along the communication path.

In the multicast case, the Controller also decides the path that the packet has to follow in order to reach all the nodes of the group.
Typically, the algorithm used to construct that path is the Steiner tree, with complexity depending on the size of the whole network.
However, when using geographic routing, the Euclidean Steiner tree algorithm can be leveraged, which has complexity dependent only on the number of nodes in the mulicast group.
Observe that the Euclidean Steiner tree algorithm may introduce Steiner (branching) points, which do not necessarily correspond to nodes of the multicast group, however they are necessary in order to optimize the routing path.
In fact, Steiner points are artificial, as there might not be a network node corresponding to their coordinates.
In this case, the node closest to these coordinates is selected as a Steiner point.
In the rest  of this section, this node will be referred to as Steiner node.

When a node wants to send a multicast packet, it sets the group address as the destination and sends a request to the Controller.
Then, Controller calculates the Euclidean Steiner tree and replies with the destination coordinates of the next multicast or Steiner node.
Nodes use geographic routing, as described above, to forward the packets towards each multicast or Steiner node.
When a multicast or Steiner node receives a packet, it sends a request to the Controller, which sends back the next multicast or Steiner node.
This process is repeated until the packet has reached all nodes in the multicast group.

The implementation of both the geographic unicast and multicast is made as an SD-WISE application.
New packet types and formats have been introduced in order to manage the geographic-related requests by the particular application.
Moreover, group management operations have also been implemented by following a protocol similar to IGMP.
Geographic operations are made available on the sensor nodes by leveraging the 
NFV capabilities enabled by SD-WISE.
In fact, in case geographic routing is enabled, SD-WISE OS sends a message to 
the nodes at system bootstrap with the function returning the intermediate node 
which is the nearest to the destination, as well as a rule to call it.
This rule is triggered when the node receives 
the coordinates of the destination and has to forward the packet to the next 
hop.
% Details on the protocol specification, implementation and packet formats can be found in~\cite{Anadiotis2017}.

\begin{figure*}
 \centering
 \subfloat[CDF of the overall number of signaling messages for different unicast forwarding strategies] {
  \label{fig:signaling}
  \includegraphics[width=0.5\textwidth]{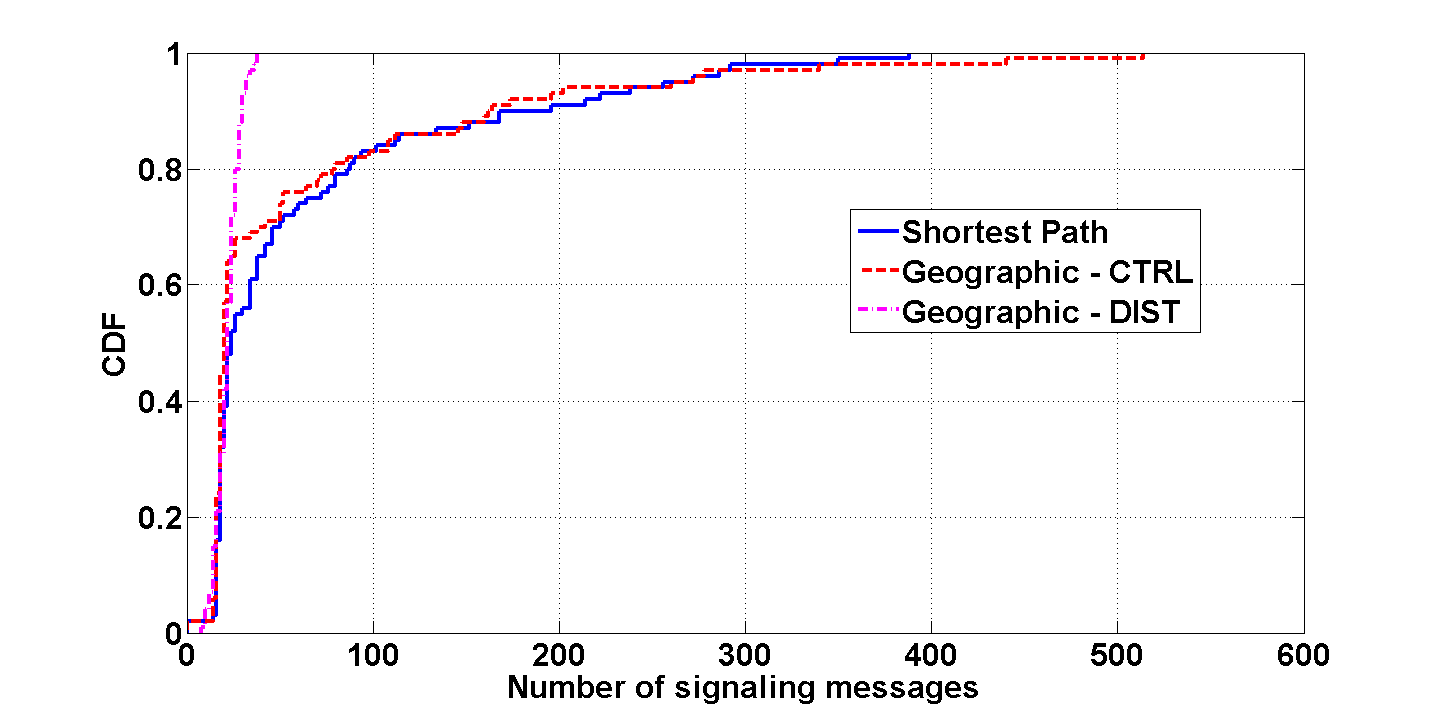}
 }\qquad
 \subfloat[CDF of the number of rules for different unicast forwarding strategies] {
  \label{fig:rules}
  \includegraphics[width=0.43\textwidth]{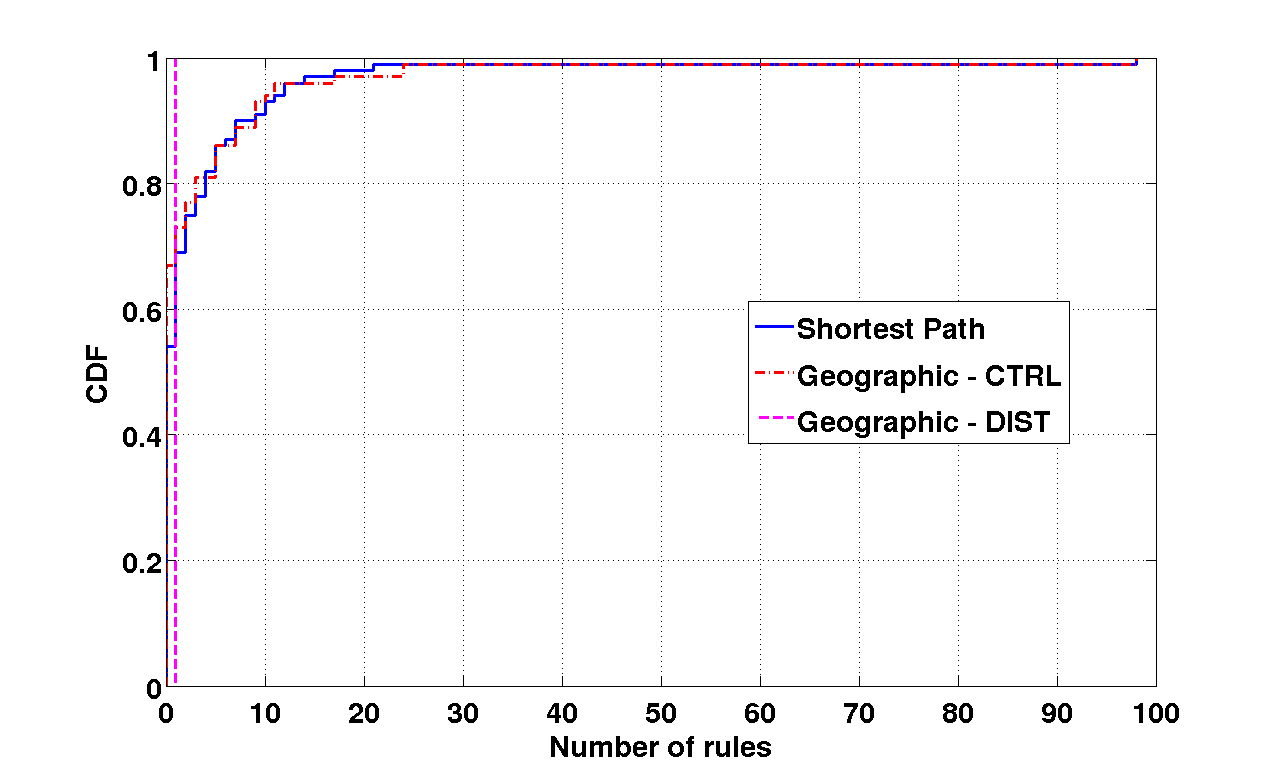}
 }%\qquad
%  \subfloat[CDF of the energy consumption in the unicast case for the considered forwarding strategies] {
%   \label{fig:energy}
%   \includegraphics[width=0.3\textwidth]{figures/energy_cdf.png}
%  }
 \caption{Impact of geographic routing on the size of flow tables and signaling}
 \label{fig:evaluation}
\end{figure*}

\begin{figure}
\centering
\includegraphics[width=\columnwidth]{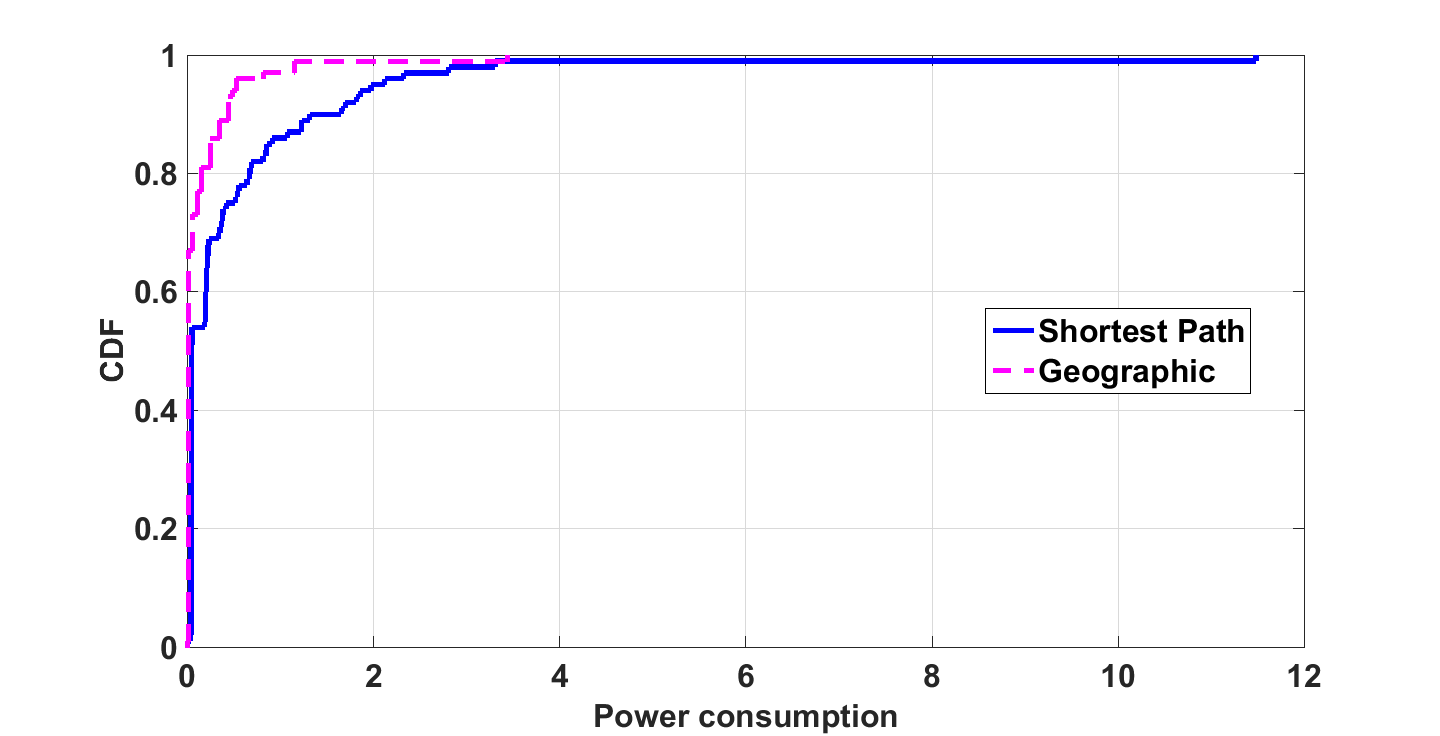}
\caption{CDF of the energy consumption in the unicast case for the considered forwarding strategies}
\label{fig:energy}
\end{figure}

The performance of geographic forwarding in SD-WISE have been evaluated using the Mininet emulator.
More specifically, we consider a 80x80$m^2$ area with 100 nodes. 
The positions of the nodes were generated randomly according to uniform distribution.
There is one sink, which acts as a gateway between the WSN and the outside world, including SD-WISE OS.
Even though we considered both the case of unicast and multicast routing, we present the results of the unicast case only, since multicast follows the same trends, as the forwarding decisions are made based on the same principles.
According to the already described protocol specification, all nodes know their own coordinates, as well as the coordinates of their neighbors, from SD-WISE OS.

We compare the following approaches: (i) \textbf{Shortest path} where the SD-WISE OS estimates the shortest path to reach the destination using the Dijkstra algorithm and sends back this information to the source node so that intermediate nodes simply relay the packet according to a pre-computed path; (ii) \textbf{Geographic-CTRL} where SD-WISE OS preliminarily decides on the geographic forwarding paths, so that intermediate nodes simply relay the packets; (iii) \textbf{Geographic-DIST} where the distributed geographic forwarding is implemented as already described earlier in this section.

Figure \ref{fig:evaluation} depicts the impact of geographic forwarding on the number of signaling messages (Fig. (\ref{fig:signaling})) and on the number of forwarding rules installed on the nodes (Fig. (\ref{fig:rules})).
As clearly shown, the Geographic-DIST forwarding case outperforms the Geographic-CTRL case, even if the latter still considers geographic routing.
The reason is that in Geographic-DIST, SD-WISE OS only needs to send the coordinates of the destination and all the other decisions are made independently by each node.
The impact of this behavior is outlined in Fig. \ref{fig:energy}, which shows the CDF of the energy consumption of the nodes.
Since most of the energy consumption of the sensor nodes is due to communication, the reduction of signaling strongly affects the energy consumption and, therefore, the overall network lifetime.

%%%%%%%%%%%%%%%%%%%%%%%%%%%%%%%%%%
\subsection{Context-based fencing} \label{ContextBasedFencing}
%%%%%%%%%%%%%%%%%%%%%%%%%%%%%%%%%%

As already described in Section \ref{RelatedWork}, there are several examples of applications where context-based regulations are required.

Given the virtually endless combination of devices, environments, and regulations, this section will describe a sample application that, although specific, will be taken as a model to generalize the requirements and design tradeoffs to be considered in similar deployments.

The proposed use case consists of a drone equipped with a camera that is allowed to record a video only if it is flying over a certain area and it is oriented towards a particular target in order to avoid copyright or privacy infringements (e.g. the drone is flying over an open air concert).
In this case, the context of the device being considered is given by the status of the camera, the position and orientation of the drone, and the status of the activity within the area framed by the camera.
All of these information are used by the Recognized Authority to choose what limitations should be imposed to the device. These limitations should have priority over the commands and configurations decided by the user of the device and, at the same time, must be implemented using  information that are up to date and trustworthy. To achieve such result, the drone is equipped with a TPM 
which guarantees that the firmware running on the device has not been tampered and the measurements coming from the GPS and accelerometers sensors mounted on the device are authentic. 

It should be emphasized that, given the limited resources in terms of storage and computation on most of the controlled devices and to allow a dynamic activation of such restrictions, the controlled devices have to report their context to their cyber counterpart.

From a communication point of view, this design requirements imposes a trade-off between the amount of information exchanged with the Recognized Authority
%virtual device
 and the delay after which these restrictions become active.

\begin{figure}[!t]
\centering
\includegraphics[width=3.2in]{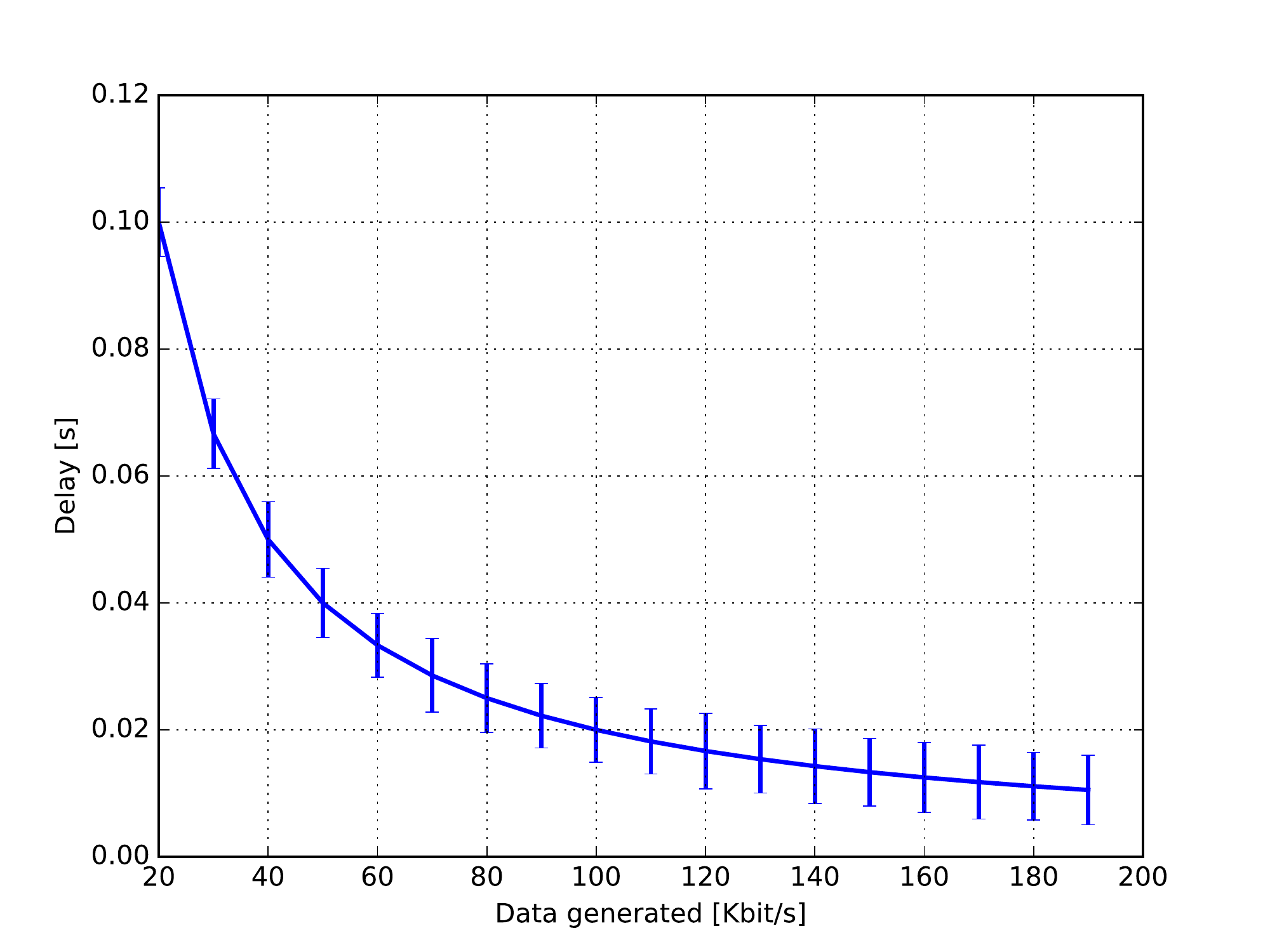}
\caption{Trade-off between data rate and activation delay.}
\label{delay}
\end{figure}

In fact, the frequency of such information exchange can be easily changed taking into account that:
\begin{itemize}
\item
The higher the chosen frequency, the higher the communication cost, which in many cases is the most relevant key performance metric;
\item
The lower the chosen frequency, the less reactive is the solution to rapid changes of context, which is critical in cases where the restriction policy dictated by the Regulation Authority depends on some parameters that change rapidly.
\end{itemize}
The trade-off between the amount of data transmitted and the average activation delay is shown in Figure \ref{delay}.

The $x$ axis shows the amount of data generated which is a function of the chosen signaling frequency and the format used to transmit the data.
On the $y$ axis, instead, the average delay after which the restriction becomes operative is reported.
The choice of the signaling frequency can be posed as a minimization problem where the cost function to be minimized is $cost = a \times rate + b \times delay$ where $a$ and $b$ are the weights chosen by the user depending on the application requirements.

%is a weighted sum of the data generated for signaling purposes and the delay, i.e.,

%%%%%%%%%%%%%%%%%%%%%%%%%%%%%%%%%%%%%%%%%%%%%%%%%%
\section{Conclusions} \label{Conclusions}
%%%%%%%%%%%%%%%%%%%%%%%%%%%%%%%%%%%%%%%%%%%%%%%%%%%%%%%%%%%%

In this paper we have introduced a Software-Defined WIreless SEnsor networking solution called SD-WISE.
SD-WISE extends the SDN approach to wireless sensor networks and introduces two major novelties when compared to similar solutions.
First of all, SD-WISE leverages existence of operating systems for wireless sensor nodes to support the network function virtualization (NFV) paradigm which can be applied to implement any networking function.
As an example, in this paper we have exploited the NFV paradigm to implement geographic routing.

Furthermore, SD-WISE exploits the strict interplay between trusted hardware and software to guarantee that sensor nodes will behave as imposed by a remote recognized authority on the basis of the current context. 
To this purpose SD-WISE leverages software and sensor attestation mechanisms supported by \emph{trusted platform modules} (TPM).
In this way SD-WISE can be considered the enabling technology of a new family of trustworthy wireless sensor networks whose behavior can be controlled to comply with context-based regulations.


%\nocite{*}
\bibliographystyle{IEEEtran}
% \bibliography{manuscript-gm}

%\bibliography{microfluidic-bibliography5_5}

\begin{thebibliography}{10}
\providecommand{\url}[1]{#1}
\csname url@samestyle\endcsname
\providecommand{\newblock}{\relax}
\providecommand{\bibinfo}[2]{#2}
\providecommand{\BIBentrySTDinterwordspacing}{\spaceskip=0pt\relax}
\providecommand{\BIBentryALTinterwordstretchfactor}{4}
\providecommand{\BIBentryALTinterwordspacing}{\spaceskip=\fontdimen2\font plus
\BIBentryALTinterwordstretchfactor\fontdimen3\font minus
  \fontdimen4\font\relax}
\providecommand{\BIBforeignlanguage}[2]{{%
\expandafter\ifx\csname l@#1\endcsname\relax
\typeout{** WARNING: IEEEtran.bst: No hyphenation pattern has been}%
\typeout{** loaded for the language `#1'. Using the pattern for}%
\typeout{** the default language instead.}%
\else
\language=\csname l@#1\endcsname
\fi
#2}}
\providecommand{\BIBdecl}{\relax}
\BIBdecl

\bibitem{Contiki}
\BIBentryALTinterwordspacing
Contiki website. [Online]. Available: \url{http://www.contiki-os.org/}
\BIBentrySTDinterwordspacing

\bibitem{RIOT}
\BIBentryALTinterwordspacing
Riot website. [Online]. Available: \url{https://riot-os.org/}
\BIBentrySTDinterwordspacing

\bibitem{6324377}
T.~Luo, H.~P. Tan, and T.~Q.~S. Quek, ``{Sensor OpenFlow: Enabling
  Software-Defined Wireless Sensor Networks},'' \emph{IEEE Communications
  Letters}, vol.~16, no.~11, pp. 1896--1899, November 2012.

\bibitem{6385039}
S.~Costanzo, L.~Galluccio, G.~Morabito, and S.~Palazzo, ``Software defined
  wireless networks: Unbridling {SDNs},'' in \emph{2012 European Workshop on
  Software Defined Networking}, Oct 2012, pp. 1--6.

\bibitem{7387950}
B.~T. de~Oliveira, L.~B. Gabriel, and C.~B. Margi, ``{TinySDN: Enabling
  Multiple Controllers for Software-Defined Wireless Sensor Networks},''
  \emph{IEEE Latin America Transactions}, vol.~13, no.~11, pp. 3690--3696, Nov
  2015.

\bibitem{7172291}
C.~Buratti, A.~Stajkic, G.~Gardasevic, S.~Milardo, M.~D. Abrignani, S.~Mijovic,
  G.~Morabito, and R.~Verdone, ``{Testing Protocols for the Internet of Things
  on the EuWIn Platform},'' \emph{IEEE Internet of Things Journal}, vol.~3,
  no.~1, pp. 124--133, Feb 2016.

\bibitem{McKeown:2008:OEI:1355734.1355746}
N.~McKeown, T.~Anderson, H.~Balakrishnan, G.~Parulkar, L.~Peterson, J.~Rexford,
  S.~Shenker, and J.~Turner, ``{{OpenFlow}: Enabling Innovation in Campus
  Networks},'' \emph{SIGCOMM Comput. Commun. Rev.}, vol.~38, no.~2, pp. 69--74,
  March 2008.

\bibitem{Casado:2006:SPA:1267336.1267346}
M.~Casado, T.~Garfinkel, A.~Akella, M.~J. Freedman, D.~Boneh, N.~McKeown, and
  S.~Shenker, ``{SANE: A Protection Architecture for Enterprise Networks},'' in
  \emph{Proceedings of the 15th Conference on USENIX Security Symposium -
  Volume 15}, ser. USENIX-SS'06, 2006.

\bibitem{Casado:2007:ETC:1282427.1282382}
M.~Casado, M.~J. Freedman, J.~Pettit, J.~Luo, N.~McKeown, and S.~Shenker,
  ``{Ethane: Taking Control of the Enterprise},'' \emph{SIGCOMM Comput. Commun.
  Rev.}, vol.~37, no.~4, pp. 1--12, Aug. 2007.

\bibitem{Greenberg:2005:CSA:1096536.1096541}
A.~Greenberg, G.~Hjalmtysson, D.~A. Maltz, A.~Myers, J.~Rexford, G.~Xie,
  H.~Yan, J.~Zhan, and H.~Zhang, ``{A Clean Slate 4D Approach to Network
  Control and Management},'' \emph{SIGCOMM Comput. Commun. Rev.}, vol.~35,
  no.~5, pp. 41--54, Oct. 2005.

\bibitem{Caesar:2005:DIR:1251203.1251205}
M.~Caesar, D.~Caldwell, N.~Feamster, J.~Rexford, A.~Shaikh, and J.~van~der
  Merwe, ``Design and implementation of a routing control platform,'' in
  \emph{Proceedings of the 2Nd Conference on Symposium on Networked Systems
  Design \& Implementation - Volume 2}, ser. NSDI'05, 2005, pp. 15--28.

\bibitem{Gude:2008:NTO:1384609.1384625}
N.~Gude, T.~Koponen, J.~Pettit, B.~Pfaff, M.~Casado, N.~McKeown, and
  S.~Shenker, ``{NOX: Towards an Operating System for Networks},''
  \emph{SIGCOMM Comput. Commun. Rev.}, vol.~38, no.~3, pp. 105--110, Jul. 2008.

\bibitem{floodlightweb}
\BIBentryALTinterwordspacing
``{Floodlight OpenFlow Controller}.'' [Online]. Available:
  \url{http://www.projectfloodlight.org/floodlight}
\BIBentrySTDinterwordspacing

\bibitem{Erickson:2013:BOC:2491185.2491189}
\BIBentryALTinterwordspacing
D.~Erickson, ``The {Beacon Openflow Controller},'' in \emph{Proceedings of the
  Second ACM SIGCOMM Workshop on Hot Topics in Software Defined Networking},
  ser. HotSDN '13.\hskip 1em plus 0.5em minus 0.4em\relax New York, NY, USA:
  ACM, 2013, pp. 13--18. [Online]. Available:
  \url{http://doi.acm.org/10.1145/2491185.2491189}
\BIBentrySTDinterwordspacing

\bibitem{6918985}
J.~Medved, R.~Varga, A.~Tkacik, and K.~Gray, ``{OpenDaylight: Towards a
  Model-Driven SDN Controller architecture},'' in \emph{Proceeding of IEEE
  International Symposium on a World of Wireless, Mobile and Multimedia
  Networks 2014}, June 2014, pp. 1--6.

\bibitem{Koponen:2010:ODC:1924943.1924968}
T.~Koponen, M.~Casado, N.~Gude, J.~Stribling, L.~Poutievski, M.~Zhu,
  R.~Ramanathan, Y.~Iwata, H.~Inoue, T.~Hama, and S.~Shenker, ``Onix: A
  distributed control platform for large-scale production networks,'' in
  \emph{Proceedings of the 9th USENIX Conference on Operating Systems Design
  and Implementation}, ser. OSDI'10, 2010, pp. 1--6.

\bibitem{Berde:2014:OTO:2620728.2620744}
\BIBentryALTinterwordspacing
P.~Berde, M.~Gerola, J.~Hart, Y.~Higuchi, M.~Kobayashi, T.~Koide, B.~Lantz,
  B.~O'Connor, P.~Radoslavov, W.~Snow, and G.~Parulkar, ``{ONOS: Towards an
  Open, Distributed SDN OS},'' in \emph{Proceedings of the Third Workshop on
  Hot Topics in Software Defined Networking}, ser. HotSDN '14.\hskip 1em plus
  0.5em minus 0.4em\relax New York, NY, USA: ACM, 2014, pp. 1--6. [Online].
  Available: \url{http://doi.acm.org/10.1145/2620728.2620744}
\BIBentrySTDinterwordspacing

\bibitem{7389118}
A.~C.~G. Anadiotis, L.~Galluccio, S.~Milardo, G.~Morabito, and S.~Palazzo,
  ``Towards a software-defined network operating system for the {IoT},'' in
  \emph{2015 IEEE 2nd World Forum on Internet of Things (WF-IoT)}, Dec 2015,
  pp. 579--584.

\bibitem{Tiscareno2016}
V.~M. Tiscareno, K.~W. Jonhson, and C.~H. Lawrence, ``Systems and methods for
  receiving infrared data with a camera designed to detect images based on
  visible light,'' in \emph{US patent 9,380,225}, June 2016.

\bibitem{Brasser:2016:RAT:2906388.2906390}
\BIBentryALTinterwordspacing
F.~Brasser, D.~Kim, C.~Liebchen, V.~Ganapathy, L.~Iftode, and A.-R. Sadeghi,
  ``Regulating arm trustzone devices in restricted spaces,'' in
  \emph{Proceedings of the 14th Annual International Conference on Mobile
  Systems, Applications, and Services}, ser. MobiSys '16.\hskip 1em plus 0.5em
  minus 0.4em\relax New York, NY, USA: ACM, 2016, pp. 413--425. [Online].
  Available: \url{http://doi.acm.org/10.1145/2906388.2906390}
\BIBentrySTDinterwordspacing

\bibitem{tpm_spec}
\BIBentryALTinterwordspacing
Tpm design principles - trusted computing group. [Online]. Available:
  \url{http://bit.ly/2sBd2rC}
\BIBentrySTDinterwordspacing

\bibitem{Truong2015}
H.-L. Truong and S.~Dustdar, ``Principles for engineering iot cloud systems,''
  \emph{IEEE Cloud Computing}, vol.~2, no.~2, pp. 68--76, 2015.

\bibitem{Schilit1994}
B.~Schilit, N.~Adams, and R.~Want, ``Context-aware computing applications,'' in
  \emph{Proc. of WMCSA 1994}, December 1994.

\bibitem{Abowd2001}
G.~D. Abowd, A.~K. Dey, P.~J. Brown, N.~Davies, M.~Smith, and P.~Steggles,
  ``Towards a better understanding of context and context-awareness,''
  \emph{Lecture Notes in Computer Science}, vol. 1707, pp. 304--307, November
  2001.

\bibitem{Liu2012}
H.~Liu, S.~Saroiu, A.~Wolman, and H.~Raj, ``Software abstractions for trusted
  sensors.'' in \emph{In Proc. of ACM Mobisys 2012}, June 2012.

\bibitem{1255648}
M.~Zorzi and R.~R. Rao, ``Geographic random forwarding (geraf) for ad hoc and
  sensor networks: multihop performance,'' \emph{IEEE Transactions on Mobile
  Computing}, vol.~2, no.~4, pp. 337--348, Oct 2003.

\bibitem{1498457}
D.~Ferrara, L.~Galluccio, A.~Leonardi, G.~Morabito, and S.~Palazzo, ``{MACRO}:
  an integrated mac/routing protocol for geographic forwarding in wireless
  sensor networks,'' in \emph{Proceedings IEEE 24th Annual Joint Conference of
  the IEEE Computer and Communications Societies ({INFOCOM}).}, vol.~3, March
  2005, pp. 1770--1781 vol. 3.

\bibitem{6407472}
L.~Galluccio, G.~Morabito, and S.~Palazzo, ``Geographic multicast (gem) for
  dense wireless networks: Protocol design and performance analysis,''
  \emph{IEEE/ACM Transactions on Networking}, vol.~21, no.~4, pp. 1332--1346,
  Aug 2013.

\bibitem{Bachrach2005}
\BIBentryALTinterwordspacing
J.~Bachrach and C.~Taylor, \emph{Localization in Sensor Networks}.\hskip 1em
  plus 0.5em minus 0.4em\relax John Wiley \& Sons, Inc., 2005, pp. 277--310.
  [Online]. Available: \url{http://dx.doi.org/10.1002/047174414X.ch9}
\BIBentrySTDinterwordspacing

\bibitem{Mao20072529}
\BIBentryALTinterwordspacing
G.~Mao, B.~Fidan, and B.~D. Anderson, ``{Wireless sensor network localization
  techniques},'' \emph{Computer Networks}, vol.~51, no.~10, pp. 2529--2553,
  2007. [Online]. Available: \url{http://bit.ly/2sVRUPi}
\BIBentrySTDinterwordspacing

\bibitem{5972367}
V.~Daiya, J.~Ebenezer, S.~A. V.~S. Murty, and B.~Raj, ``Experimental analysis
  of rssi for distance and position estimation,'' in \emph{2011 International
  Conference on Recent Trends in Information Technology (ICRTIT)}, June 2011,
  pp. 1093--1098.

\end{thebibliography}
\end{document}